\documentclass{aa}  

\usepackage{graphicx}
\usepackage{txfonts}

\newcommand{\teff}{$T_{\rm eff}$}
\newcommand{\logg}{$\log\,{g}$}

\begin{document} 

  \title{Yellow hypergiant V509\,Cas -- stable in the 'yellow void'}

  \author{A. Kasikov\inst{1}\inst{2}\inst{3}\and
          I. Kolka\inst{1}\and
          A. Aret\inst{1}\and
          T. Eenm\"ae\inst{1}\and 
          V. Checha\inst{1}
          }

  \institute{Tartu Observatory, University of Tartu, Observatooriumi 1, Tõravere, 61602, Estonia\\
              \email{anni.kasikov@ut.ee}
        \and
            Nordic Optical Telescope, Roque de Los Muchachos Observatory, Rambla Jose Ana Fernandez Perez 7, 38711 Brena Baja, La Palma, Canarias, Spain
        \and
            Department of Physics and Astronomy, Aarhus University, NyMunkegade 120, DK-8000 Aarhus C, Denmark
            }

  \date{Recieved ; accepted }

  \abstract
  {The yellow hypergiant star \object{V509\,Cas} is currently undergoing an extreme phase of evolution. Having experienced eruptive mass-loss outbursts in the 20th century, the star's effective temperature reached record high values in the early 2000s. However, since then, the star's behaviour has displayed an unprecedented level of stability. In spite of that, the star could be traversing through the 'yellow void' instability region. 
  }
  {To describe the current evolutionary state of V509\,Cas, we analysed its variability using photometric and spectroscopic data collected over recent years. By comparing our findings with historical records, we aim to determine whether the star's surface shows signs of stabilisation. Additionally, we investigate the variability of emission components in the wings of certain spectral lines to highlight the contribution of the  circumstellar gaseous disc to this phenomenon.}
  {  Our spectroscopic monitoring observations were carried out at Tartu Observatory over the course of seven years, supplemented by echelle spectra obtained at the Nordic Optical Telescope, as well as publicly available photometric data from Gaia, AAVSO, and AAVSO's Bright Star Monitor programme. We estimated the variability of effective temperature and radial velocity from the spectral time series and correlated it with the brightness variability of V509\,Cas. 
  }
  {The results indicate that the star's average brightness level has remained stable throughout the observed period, with an amplitude of variability $ \sim$ 0.1 mag. While the amplitude of short-term temperature fluctuations has decreased compared to the early 2000s, the variability of the radial velocity remains similar to historical values from the early 20th century. Moreover, we show how the variable radial velocity affects the emission components in some absorption lines (e.g. \ion{Sc}{II}) and how that follows the hypothesis of a disc surrounding the star.} 
  {}


  \maketitle

\section{Introduction}
Yellow hypergiant stars (YHGs) are objects of great interest. They are in an extreme transitional phase of evolution, characterised by dynamically unstable atmospheres and recurring mass-loss outbursts. YHGs form a potential bridge in the evolutionary gap between red supergiant (RSG) and B[e] supergiant or luminous blue variable (LBV) phases.
    
The YHGs are positioned at the very upper part of the Hertzsprung-Russell (HR) diagram. They have spectral classes F-G and an extremely high luminosity $\left(\log L/L_\odot = 5\ldots6\right)$ \citep{Nieuwenhuijzen2012}, corresponding to an Ia+ luminosity class. The initial zero-age main sequence (ZAMS) masses are $\sim 20-40\,M_\odot$. YHGs are considered to be post-red-evolved massive stars whose temperature started to increase after their red supergiant phase. Now they are on a 'blue loop' at the very upper part of the HR diagram, where their high luminosity comes from burning helium in the core. The area in the HR diagram where we find YHGs is relatively sparsely populated and has been referred to as the 'yellow evolutionary void' (or yellow void in short) by \cite{deJager1998}. This region signifies an instability area for blueward evolving stars, and the YHGs cluster near its lower temperature boundary. 
    
Stars in the yellow void region have extended atmospheres and experience high rates of mass loss due to their low or near-zero surface gravity. YHGs undergo eruptive episodes lasting about one to two years \citep{vanGenderen2019}, during which the upper atmospheric layers are expelled, resulting in a cooler envelope forming around the star. This leads to a drop in both the star's brightness and effective temperature. Subsequently, the star's temperature gradually rises until another eruption occurs. Such episodes can repeat multiple times and a large fraction of the star's mass may be ejected \citep{Nieuwenhuijzen2000,Nieuwenhuijzen2012}. The current masses of YHGs are between $15-25\,M_\odot$ \citep{Nieu1995}. Eventually the star could reach a more stable phase \citep{deJager1998} and transition out of the YHG phase and emerge on the other side of the void as an evolved blue supergiant -- B[e] or LBV-type star \citep{Aret2017}. The yellow hypergiant phase is very short-lived, lasting less than $10^5$ years, which explains the small number of known hypergiants \citep{deJager1998}.
    
V509\,Cas is a representative member of the YHGs and has experienced multiple large-scale shell ejections. \citet{Smolinski1989} described three episodes of mass ejections: in 1970, 1979, and 1982. Changes in the star's effective temperature (\teff) can be linked to these mass-loss episodes. When the temperature of V509\,Cas increased, it then approached the low-temperature border of the yellow void in the HR diagram. This resulted in enhanced mass loss due to which an optically thick shell formed around the star. The cooler shell caused a decrease in the star's \teff. Following the episode, the shell dispersed and the star's {\teff} seemed to increase again. In the HR diagram, the behaviour of V509\,Cas can be described as 'bouncing' against the low-temperature border of the yellow void \citep{Nieuwenhuijzen2000}. There is no indication of distant nebulosity around V509\,Cas, suggesting that if the star is a post-RSG, it has only recently (within the last 500--1000 years) entered the instability region \citep{Schuster2006}. 
    
A more recent extensive study of the yellow hypergiant V509\,Cas has been conducted by \citet{Nieuwenhuijzen2012}, providing a comprehensive overview of various physical parameters of the star, including temperature, luminosity, colour indices, and \logg\ values. They also challenge the definition of the yellow evolutionary void and propose that it could consist of two separate instability regions. In the cooler first part ($3.8 < \log T_{\rm eff} < 3.95$), the primary factor driving instability is the ionisation of hydrogen and in the hotter second part ($4.05 < \log T_{\rm eff} < 4.15$) ionisation of helium dominates. It is expected that as the star approaches the high temperature border of the first instability region, it becomes more stable. However, upon crossing over to the second area of instability, the atmosphere may become unstable again, leading to severe mass loss. \cite{Aret2017} support the suggestion of \cite{Nieuwenhuijzen2012} that V509\,Cas is approaching the high temperature boundary of the first instability region in the yellow void. 

The link between the observed variability and the evolution of YHGs remains a subject of debate. The evolution of a star in the yellow void could be rapid, occurring over a timescale of a few thousand years \citep{Lobel2017}. Alternatively, it is possible that only the outer layers of the star's atmosphere are affected during the enhanced mass-loss episodes as the hypergiant bounces against the low-temperature border of the yellow void, while the evolution of the star underneath continues uninterrupted by apparent eruptive changes \citep{deJager1997}. 

The main trigger for mass loss appears to be pulsations \citep{deJager1998}. The pulsations are quasi-periodic and they cause variability in the star's brightness, with an amplitude ranging from 0.2 to 0.5~mag. The variable periods and amplitudes in the light curve depend on the effective temperature of the star -- when the star is in a hotter state, the periods are shorter and amplitudes are smaller; when the star is cooler, the periods are longer and amplitudes are larger \citep{vanGenderen2019}. 

Large-scale nebulosity has not been detected around V509 Cas \citep{Schuster2003}, but not much is known of the small-scale environment of V509\,Cas. \cite{Aret2017} suggested the presence of an inner disc based on double-peaked [\ion{Ca}{II}] lines and a single-peaked [\ion{O}{I}] $\lambda 6300$ line. As the star loses mass, the expelled material could accumulate in the equatorial plane, potentially giving rise to a disc-like structure. The presence of a disc around the star might suggest an interesting evolutionary path, potentially leading to the formation of a massive disc, such as those found around B[e] SGs.

Currently, V509\,Cas is located in the HR diagram at the low-temperature boundary of the yellow void, thus providing a strong rationale for close monitoring of this target. Over the past seven years, consistent spectral monitoring of V509\,Cas has been conducted at Tartu Observatory\footnote{\url{https://kosmos.ut.ee/}} (TO), resulting in a substantial dataset of spectra that can be used to characterise the star's behaviour at the border of instability. Additionally, several high-resolution spectra have been obtained at the Nordic Optical Telescope\footnote{\url{http://www.not.iac.es/}} (NOT). Photometric data from Gaia \citep{Gaiamission, GaiaDR3}, TESS\footnote{\url{https://archive.stsci.edu/missions-and-data/tess}} and the American Association of Variable Star Observers\footnote{\url{https://www.aavso.org}} (AAVSO) and AAVSO's Bright Star Monitor (BSM) programme\footnote{\url{https://www.aavso.org/bsm}} have been included in the analysis. 

In this paper, we present an observational overview of V509\,Cas over recent years. Our aim is to analyse the spectral and photometric variability of the star to gain insights into its ongoing evolutionary journey. By comparing the recent data with historical records, we aim to determine whether the star's surface has begun to stabilise, indicating its passage through the yellow void. Furthermore, inspired by the work of \citet{Aret2017}, we investigate the impact of a potential disc-like structure on variable spectral lines with emission components, specifically examining a \ion{Sc}{II} line.

\section{Observations} 

\subsection{Photometry}

\subsubsection{AAVSO international database}

The majority of photometric data are obtained from the AAVSO international database. We use observations from three observers who have carried out longest time series or have transformed their data to Johnson-Cousins photometric system: W. Vollmann (AAVSO observer code VOL, observations in $V$-filter \citep{VollmannAAVSO}), M. Sblewski (SMAI, observations in $B$ filter \citep{SblewskiAAVSO}) and E. Van Ballegoij (BVE, observations in $B$-filter \citep{BallegoijAAVSO}). The observations are summarised in Table~\ref{tab:aavso_internatioan}.

\begin{table}[]
    \centering
    \begin{tabular}{l c c c c}
    \hline \hline
    Observer    & Code & Filters         & Obs. type & Data points \\ \hline
    Vollmann & VOL  & $V$   & DSLR     & 445 \\
    Sblewski & SMAI & $B$*    & CCD      & 141 \\
    Van Ballegoij & BVE & $B$ & PEP      & 54 \\ \hline
    \end{tabular}
    \caption{Observations from the AAVSO international database. The observation type refers to the used photometric measurement device, where DSLR means the digital single lens reflex camera, CCD is a CCD camera and PEP refers to measurements done with a single-channel photometer using a photodiode. The asterisk marks observations that have not been transformed to the standard Johnson-Cousins system.}
    \label{tab:aavso_internatioan}
\end{table}

\subsubsection{Observations from AAVSO Bright Star Monitor}\label{BSMobs}

The recent photometric variability of V509\,Cas is captured  by observations of AAVSO BSM program proposal \#115. The observations were initiated by T. Eenmäe.
We used Johnson-Cousins $B$- and $V$-filter observations from two observatories: New Hampshire Henden (NH2) and New Mexico Stein (NM). Both sites utilise Takahashi E180 7'' reflectors with ASI183MM CMOS-cameras and a selection of photometric filters.  
Photometric monitoring using BSM started in September 2019 to support TESS observations in sectors 16 and 17, and has continued till present time \citep{EenmaeAAVSO}.

The photometric observations from AAVSO BSM programme have been optimised for measuring bright stars. Nevertheless, the data are still affected by the scintillation effect. To mitigate these scintillation effects when using short exposures, several individual measured data frames are averaged. 
To better estimate the differential magnitude of the variable star, we used the combined comparison star method. The combined comparison star is composed of the summed flux of three stars in the field of view: \object{HD~218010}, \object{HD~240170} and \object{HD~217127}. The first two are included in the list of AAVSO comparison stars and the constant magnitude of the third comparison star has been confirmed by measurements done by the Hipparcos mission ($1\sigma$ scatter of measurements was $\pm 0.012$ mag \citep{ESAHipp}).

We obtained pre-processed frames from the BSM programme and measured the magnitudes of the stars using aperture photometry with programme Aperture Photometry Tool (APT, \citet{APT}). The aperture size was selected to be $3\times\textrm{FWHM}$ of the target. 
 We determined the differential magnitude of V509 Cas either from the average frame of a night (provided by the AAVSO BSM pipeline) or from many individual frames taken during the same night, the results of which were then averaged. 
Therefore, one point on the graph represents the luminosity of the star in one night. The number of averaged frames and their exposure times are shown in Table~\ref{table:photframes}.

\begin{table}
\centering
\begin{tabular}{l c c c}
\hline\hline
Filter & Nr of frames & Exposure & \# nights\\ 
\hline
    $B$       & 8..10 & 1.5 s & 272\\
    $V$       & 15..20 & 0.75 s & 265\\
\hline
\end{tabular}
\caption{Observations with the AAVSO BSM telescopes, filters, the number of frames taken in each filter per night, and the exposure times of each individual frame are given.} 
\label{table:photframes}
\end{table}

The results slightly vary depending on the weather conditions (airmass, transparency of the atmosphere, etc.) and scintillation. The typical root-mean-square (RMS) value is around 3\% for both V509\,Cas and the combined comparison star. We left out nights on which the conditions were highly unstable and the RMS value was found to be higher than 6\%. The resulting differential magnitude of the star (average per night) has the uncertainty $0.01\ldots0.02$~mag.

\subsubsection{TESS}\label{section:tess}

V509\,Cas was observed with the TESS satellite in sectors 16, 17, 24, and 57. The observations were performed at the following dates: 50 days in Sept-Nov 2019, 30 days in April-May 2020 and 29 days in October 2022. Although TESS does not provide luminosity measurements in the $V$-filter, these observations show the brightness variability of the star with very good cadence and precision. We measured the differential TESS magnitude of V509\,Cas against the comparison star HD~217127, both of them imaged on the same TESS camera and CCD. In the selection of proper apertures and in the correction of the light curve systematics, we have been guided by the papers by \citet{Handberg2021} and \citet{Lund2021}. The resulting magnitudes have a precision better than $\pm0.005$~mag. Therefore, we have used this data to qualitatively describe the variability of V509\,Cas and to complement the $V$-filter observations from AAVSO. 

\subsubsection{Gaia}\label{section:gaia}

We are using data from the Gaia Data Release~3 \citep{Gaiamission, GaiaDR3}. Photometric observations were done between 25th July 2014 and 28th May 2017, spanning a period of 34 months. 
Gaia has measured the brightness of V509\,Cas in three filters: $G$, $G_\textrm{BP}$, and $G_\textrm{RP}$. 
We converted those observations to the Johnson-Cousins system $V$ filter brightnesses, using the colour transformation equations given 
in the Gaia Data Release\,3 Documentation version 1.3
on the Gaia webpage \footnote{\url{https://gea.esac.esa.int/archive/documentation/GDR3/Data_processing/chap_cu5pho/cu5pho_sec_photSystem/cu5pho_ssec_photRelations.html}}. Gaia's brightness measurements have small errors (<0.001 mag for $G_\textrm{BP}$ and $G_\textrm{RP}$, 0.02 mag for $G$-filter). The more significant errors in the Gaia data come from the conversion of Gaia's filters to the Johnson-Cousins $V$-filter magnitude; the standard deviation of the colour transformation equation is 0.03 mag.

\subsection{Spectroscopy}\label{section:spectroscopy}

\subsubsection{Tartu Observatory}

V509\,Cas was observed using the Tartu Observatory 1.5-m telescope AZT-12 \citep{Folsom2022} during the period from 2015 to 2021, in total on 103 nights, fairly evenly distributed across the years. We used the long-slit spectrograph ASP-32 in Cassegrain focus with 1800\,lines mm$^{-1}$ diffraction grating providing spectra in the wavelength range from $6300\,$\AA\ to $6730\,$\AA\ with a signal-to-noise ratio (S/N) $\sim 200-300$ and resolution $R \approx 10\,000$.

The data were reduced using the {\sc iraf}\footnote{\url{https://iraf.net/}}\citep{IRAF} software, in which \texttt{noao}, \texttt{imred}, \texttt{ccdred}, and \texttt{ctioslit} packages for image and long-slit spectral reductions were used. Bias and flat corrections were applied to the spectra. Dark correction was not necessary due to the sufficient cooling of the CCD. For the wavelength calibration, comparison spectra of ThAr hollow cathode lamp were taken before and after target exposure. In addition, the wavelengths of the diffuse interstellar band (DIB) at $\lambda6379.01$ and some telluric lines near 6400~\AA\ were used to check and refine the accuracy of the wavelength calibration. The heliocentric correction was applied to the spectra. The spectra were continuum normalised by fitting a cubic spline of the 20th order. Flux calibration was not attempted. The full list of observations can be found in the appendix Table~\ref{table:TOobs}.

\subsubsection{Nordic Optical Telescope}

In addition to the observations at the TO, in 2021 and 2022 for a period of 14 months, V509\,Cas was observed at the 2.56~m Nordic Optical Telescope (NOT) using the FIES echelle spectrograph (instrument description: \citet{Telting2014}) in medium-resolution mode ($R = 45\,000$) and high-resolution mode ($R = 60\,000$) (proposal ID 65-410, PI A. Kasikov). We also include an earlier observation in 2015 in high-resolution mode ($R = 60\,000$). The spectra cover the wavelength range from $3630\,$\AA\ to $8980\,$\AA\ without gaps. 
Our observations were carried out with an approximately monthly interval, in total on 16 nights (see Table~\ref{table:FIESobs}) and were reduced with FIEStool \citep{FIEStool} standard pipeline.

\section{Analysis and results}

\subsection{Photometric variability}

V509\,Cas was not included in catalogues until the 1850s. One of the likely reasons is that it was fainter than the 6th~mag. Since then, the star has gradually increased in brightness. During the 20th century, its brightness ranged from around 5.36~mag in the beginning of the 1940s to 4.8~mag by the 1980s. The peak magnitude was reached in 1976 at 4.61~mag \citep{Zsoldos1986}. \cite{Percy1992} show the light curve of V509\,Cas from the 1970s to the mid-1990s, when the average brightness of V509\,Cas decreased from its maximum to 5.1~mag by the beginning of the 1990s. At the same time, the $B-V$ colour of the star became bluer by about 0.5~mag. After the mass-loss episode in the beginning of the 1970s, the brightness varied with an amplitude between 0.1--0.2~mag, which decreased by 1993 to about 0.05--0.1~mag. In the late 1990s and early 2000s, the variability in the Johnson $V$ filter remained in the same range \citep{vanGenderen2019}.

\subsubsection{Brightness in the $V$ filter}

\begin{figure}
    \resizebox{\hsize}{!}{\includegraphics{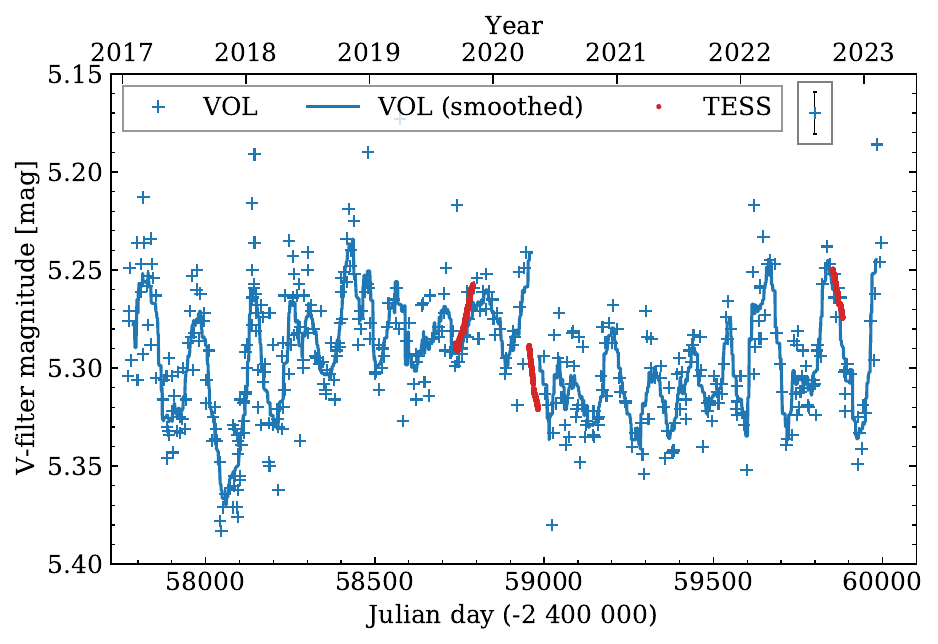}}
    \caption{Smoothing of AAVSO data taken by observer Vollmann (VOL). Shifted brightness observations from TESS (red dots) follow the smoothed curve. The mean residual of the smoothing is 0.013~mag. The mean error of observations is 0.01 mag.}
    \label{fig:aavso_smoothing}
\end{figure}

The $V$-filter observations by AAVSO (W. Vollmann) span the time period from 2017 to 2023, and they are typically conducted more frequently than once in a fortnight, with a minor gap occurring in early 2020 (see Fig.~\ref{fig:aavso_smoothing}). The mean $V$-filter magnitude has remained at a stable level throughout the observed period. Typical uncertainties of these observations average around 0.01~mag.  The brightness fluctuates without obvious regularities in the range of 5.4 to 5.2~mag. A smoothing procedure is advantageous for presenting this variability more clearly. Given the frequent sampling, we applied a straightforward 30-day weighted running mean, assigning less weight to points with higher uncertainties. 
Smoothing of $V$-filter observation data is justified by smooth brightness variability evident in the TESS light curve. While TESS did not measure brightness in the $V$-filter, our smoothed light curve effectively tracks the variability behaviour observed in TESS measurements 
(Sect. \ref{section:tess}). In Fig. \ref{fig:aavso_smoothing}, we added an arbitrary vertical shift to TESS measurements to achieve the best alignment with AAVSO data, only to illustrate the smoothness of variability.

\begin{figure}
    \resizebox{\hsize}{!}{\includegraphics{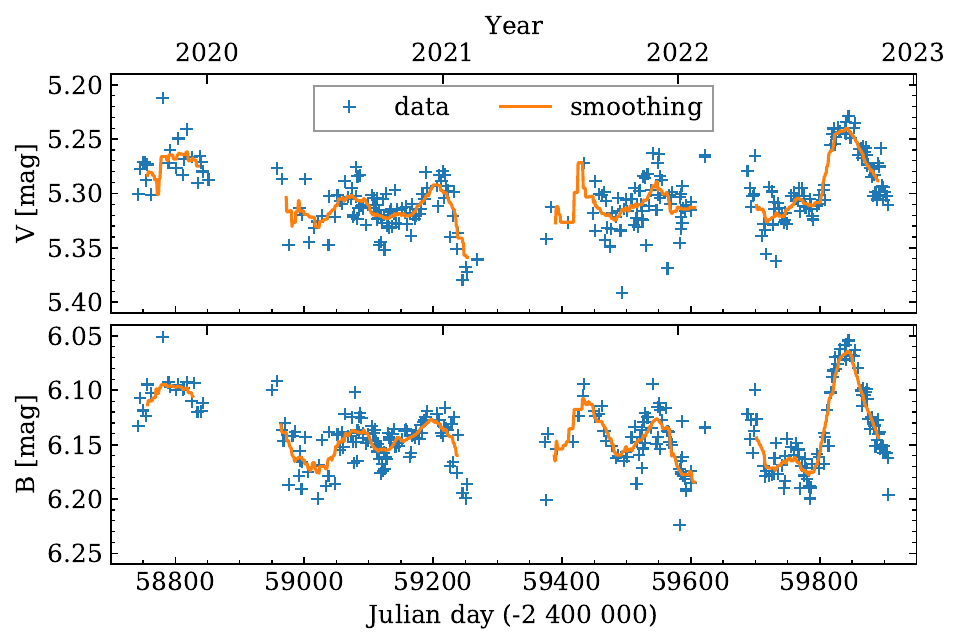}}
    \caption{Smoothing of BSM $V$- and $B$-filter data. The mean residual of the $V$-filter fit is 0.012 and the $B$-filter fit is 0.010. The uncertainty of each point is around 0.01..0.02 mag}
    \label{fig:bsm_smoothing}
\end{figure}

\begin{figure}[h]
    \resizebox{\hsize}{!}{\includegraphics{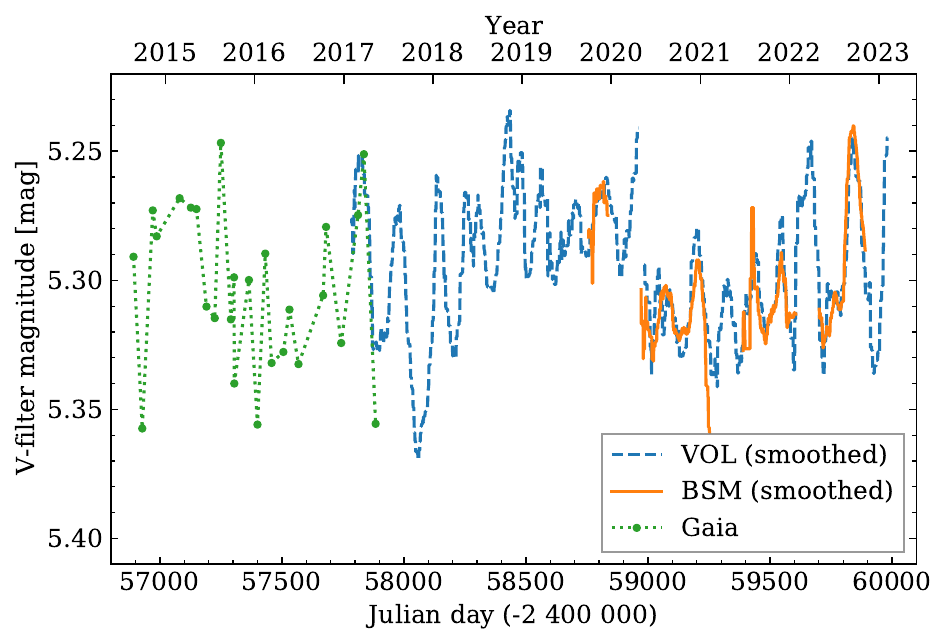}}
    \caption{Full $V$-filter light curve including data from AAVSO (VOL, smoothed), BSM and Gaia.
    }
    \label{fig:photometry}
\end{figure}

We applied the same smoothing method to BSM $B$- and $V$-filter observations -- a running mean with a 30-day window (Fig.~\ref{fig:bsm_smoothing}). The BSM magnitudes have been measured with reference to the standard stars (Sect.~\ref{BSMobs}) obtaining differential magnitudes that have not been transformed to the Johnson-Cousins standard magnitude scale. To align the observations with data provided by AAVSO observers, we shifted them by an empirical constant, which minimised the difference between two sets of data. The BSM observations show similar variability behaviour to the observations from AAVSO. We have also included observations from Gaia (Sect.~\ref{section:gaia}) to extend the time span covered by the light curve (see Fig.~\ref{fig:photometry}). 

Since the late 1990s, the brightness of V509\,Cas has stabilised. Between 2014 and 2023, it has consistently remained near 5.3 mag, displaying occasional short-term fluctuations with amplitudes up to 0.13 mag. Over the past two decades, the $V$-filter brightness variability has remained notably low, especially when compared to the rapid changes observed in the 20th century. In comparison to the Hipparcos measurements in the early 1990s (as reported by \cite{Nieuwenhuijzen2000}), the amplitudes of short-term variability have remained at a similar level.

\subsubsection{$B-V$ colour index}

\begin{figure}
    \resizebox{\hsize}{!}{\includegraphics{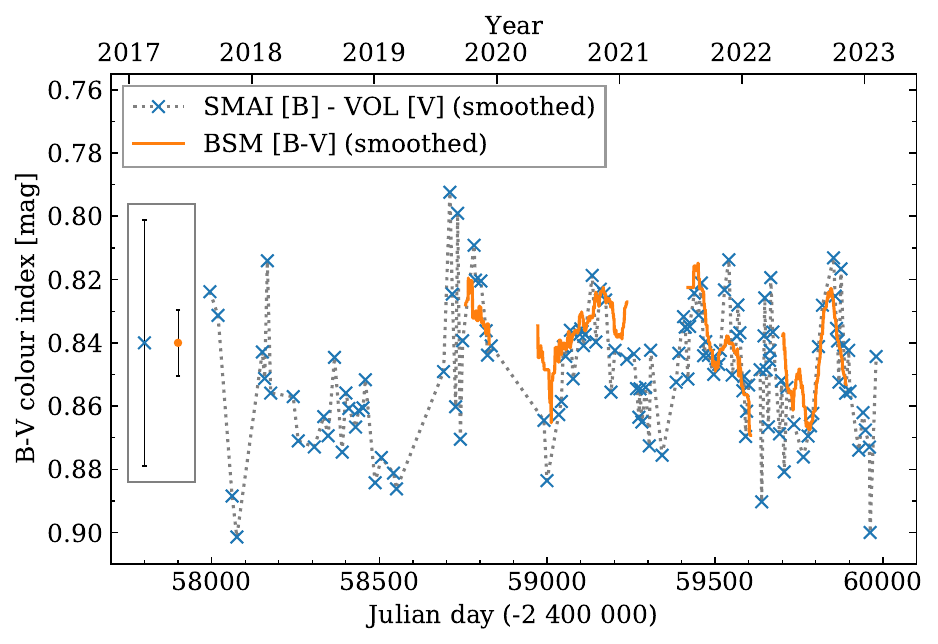}}
    \caption{$B-V$ colour index based on AAVSO observers and the BSM programme. The error bars are in the box on the left.}
    \label{fig:aavso_b-v}
\end{figure}

For the $B$-filter brightness, we have data from AAVSO observers M.~Sblewski (SMAI) and E.~Van~Ballegoij (BVE) as well as the BSM programme. The observations of Van Ballegoij have been transformed to the standard Johnson-Cousins system magnitudes, whereas Sblewski's observations have not. To align Sblewski's data with Ballegoij's, we applied a minor shift of approximately $\sim$0.02 mag, which falls well within uncertainty limits. Consequently, we made a similar shift to the BSM programme's $B$-filter measurements, allowing us to calculate the $B-V$ colour indices from the BSM programme and AAVSO. The resulting $B-V$ curves are shown in Fig.~\ref{fig:aavso_b-v}, where AAVSO and BSM colour indices exhibit very similar behaviour.

During the 1990s, when Hipparcos measured V509\,Cas, a notable decrease in $(B-V)$ values was observed \citep{Nieuwenhuijzen2012}. Based on the current data, it is apparent that the decline in $(B-V)$ values has halted. The $(B-V)$ colour of the star has remained stable over the last six years, averaging around 0.84~mag with minor short-term fluctuations in the range of 0.05--0.10~mag. As have been noted by \citet{vanGenderen2019}, amplitude ratios Ampl\,$B$/Ampl\,$V$ and Ampl\,$V$/Ampl\,$(B-V)$ reveal temperature changes during pulsations, with higher values indicating unstable atmospheric conditions. These amplitudes are independent of interstellar reddening and extinction. According to our current data, the variability amplitudes in the $V$-filter and $B$-filter are very similar, with smaller amplitudes (0.05~mag) during some years  alternating with slightly larger amplitudes ($\sim$0.10~mag) in the following few years. Currently, based on our data, both amplitude ratios have small values near 1, which is indicative of the atmosphere stabilising.

\begin{figure}
    \resizebox{\hsize}{!}{\includegraphics{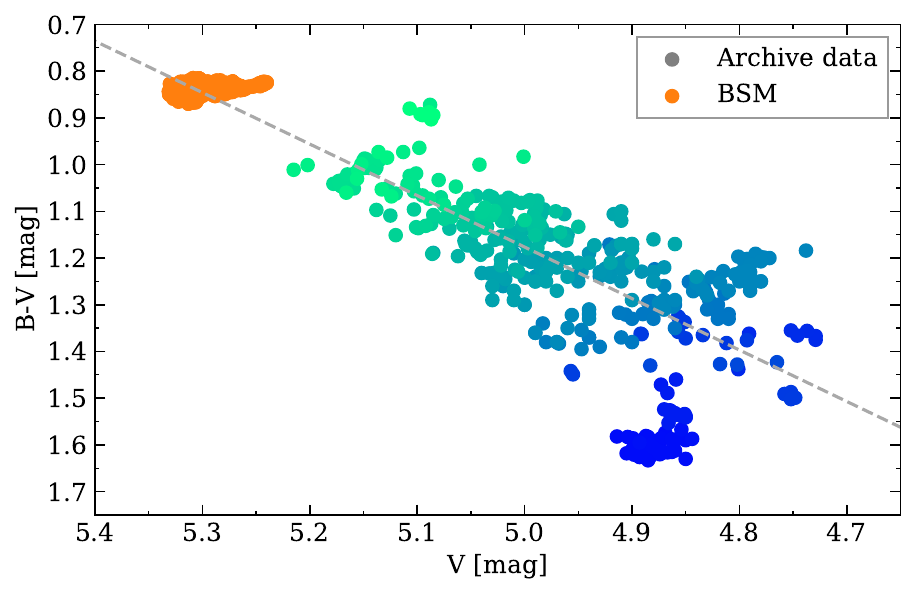}}
    \caption{Long-term correlation of the $V$-filter brightness and $B-V$ colour index. The archival data is from the paper by \citet{Nieuwenhuijzen2012} and cover the period 1976-1993. The blue-green colour gradient indicates the date of measurements: the oldest points from 1976 are dark blue and the latest points are light green. We have added the data from BSM to show the continuing trend. }
    \label{fig:v_vs_b-v}
\end{figure}

Figure~\ref{fig:v_vs_b-v} illustrates the long-term changes in the $V$-filter brightness and $B-V$ colour index. In addition to the data from the BSM programme, we have included data from \citet{Nieuwenhuijzen2012}, to provide an overview of the variability since 1976. The $(B-V)$ colour index has notably shifted towards the blue by approximately 0.8 mag, while the $V$-filter brightness has decreased by roughly 0.7 mag. The $(B-V)$ and $V$ values from AAVSO observers overlap with those from BSM, and for the sake of clarity, they have not been included in the figure.

\subsection{Periodicity of the light curve}

V509\,Cas exhibits quasi-periodic variability that is composed of several periods with varying lengths. 
A light curve compiled by \citet{Percy1992} shows new periods emerging and older ones disappearing. Additionally, the lengths and amplitudes of the periodic fluctuations change over time. They isolated three periods ranging from 200 to 400 days. Notably, periods close to 1 year have been frequently cited (e.g. \cite{Percy1992,Zsoldos1986,Sheffer1987,Nieuwenhuijzen2000}). Between 1976 and 1993 the quasi-periods of V509\,Cas decreased linearly to about 100--150 days, which was linked to the contraction of the star and increase in its atmospheric density \citep{vanGenderen2019}.

We applied the Lomb-Scargle period analysis model to the AAVSO $V$-filter data. Most of the detected periods in the AAVSO data fall within the 100--200 day range. However, there also appears to be a much longer periodicity in the data, with a period of approximately 1300 days. Neither our analysis nor earlier studies have identified this extended periodicity in the light curve from 1970s--1990s by \citet{Percy1992}.

\begin{figure}
    \resizebox{\hsize}{!}{\includegraphics{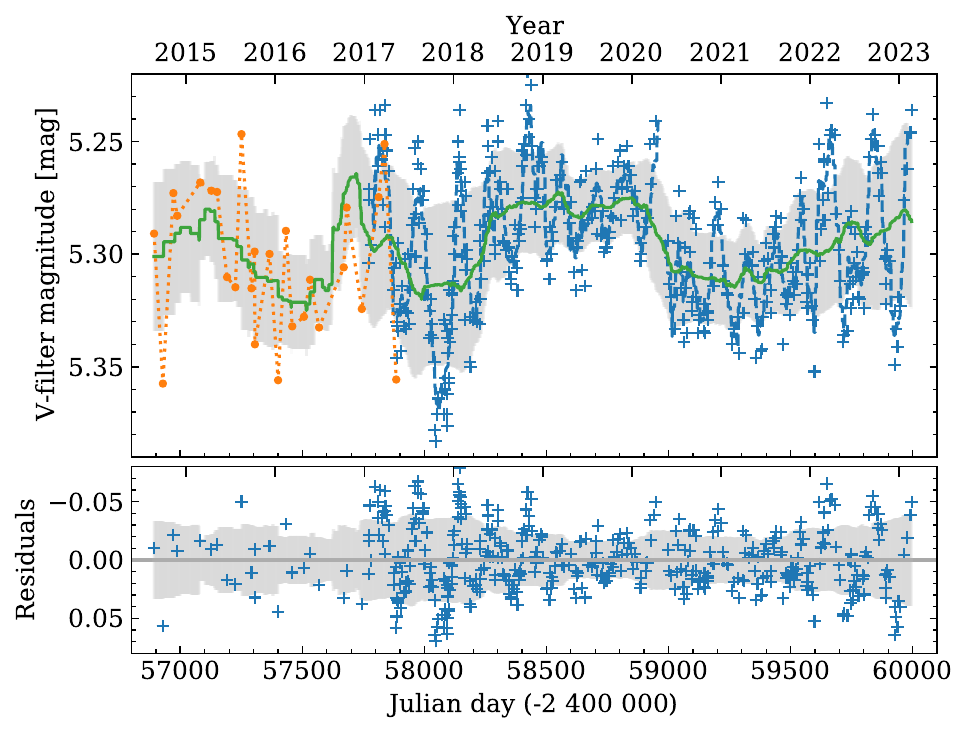}}
    \caption{$V$-filter brightness from AAVSO Vollmann's data and Gaia data. The solid green line shows a running mean of 300 days, and the light grey shading corresponds to the standard deviation of the running mean. In the lower figure we show the residuals of the mean level that illustrate the amplitudes of short-term fluctuations, which vary between 0.05--0.13 mag.}
    \label{fig:vmag_longterm}
\end{figure}

Over the past 8 years, the amplitudes of short-term brightness fluctuations in $V$-filter have varied between 0.03 and 0.13~mag. 
In Fig.~\ref{fig:vmag_longterm}, a 300-day running mean has been applied to the light curve, which smooths out short-term (\mbox{$<300$}~days) variability fluctuations and highlights variability on longer timescales. There is no evident long-term trend of decreasing or increasing variability amplitudes or mean magnitude. The variability amplitudes increase for a few years (e.g. 2017--2018) and then decrease for a few subsequent years (e.g. 2019--2020). 
The residuals in the lower panel of Fig.~\ref{fig:vmag_longterm} illustrate the variability of short-term brightness fluctuation amplitudes, when the longer-term curve has been subtracted. A longer timeline of frequent V-filter magnitude measurements would give more insight into any periodicity in the increase and decrease of amplitudes. 

\subsection{Effective temperature}

\subsubsection{Historical temperature variability}

The earliest known spectral classification of V509\,Cas dates back to the Henry Draper Catalogue (1918--1924), where it was categorised as G0p \citep{HDcat}. In 1935, its spectral class was given as G3 \citep{Adams1935}. These classes correspond roughly to temperatures of 5600~K and 5100~K, respectively, based on the classification of supergiants in Table 9 by \cite{Currie2010}. 

Over the course of the 20th century, the temperature of V509\,Cas varied by several thousand degrees. Figure~\ref{fig:teff_historic} illustrates temperature values from \cite{Nieuwenhuijzen2000}, \cite{Nieuwenhuijzen2012}, and the current study. During this period, two enhanced mass-loss episodes occurred:
the first in the early 1970s and the second in 1979--1982 \citep{deJager1998}. \citet{Israelian1999} found that during the first outburst in 1969, the temperature ranged from 5250~K to 5630~K (based on different atmosphere models). In the subsequent couple of years, during the mass-loss event, the star's temperature decreased down to 4900 K in 1973 based on analysis of \citet{Nieuwenhuijzen2000}. This decrease is attributed to the formation of a cooler pseudo-photosphere from the outflowing matter. Over time, as this pseudo-photosphere dispersed, the star's temperature began to rise again. A similar temperature pattern was observed during the second mass-loss event in 1979--1982. This temperature variability appears on the Hertzsprung-Russell diagram as a 'bouncing' motion near the yellow void atmospheric instability region. Over a span of 30 years, the star appeared to 'bounce' twice against the border of the yellow void \citep{deJager1998}. 

After these mass-loss events, the temperature of V509\,Cas continued to rise from 4570~K in 1980 \citep{Nieuwenhuijzen2000, Piters1988} to 7170~K in 1995 \citep{Nieuwenhuijzen2000}. After it reached higher-than-ever values at 7900 ($\pm$200)~K in 1998 \citep{Israelian1999}, the rate of temperature increase seems to have slowed down. In the early 21st century, the star's temperature has remained close to 8000~K \citep{Nieuwenhuijzen2012, Yamamuro2007, Aret2017}. Between 2000 and 2005, the temperature exhibited short-term variability of up to a few hundred Kelvin \citep{Nieuwenhuijzen2012}. 

\begin{figure}[h]
    \resizebox{\hsize}{!}{\includegraphics{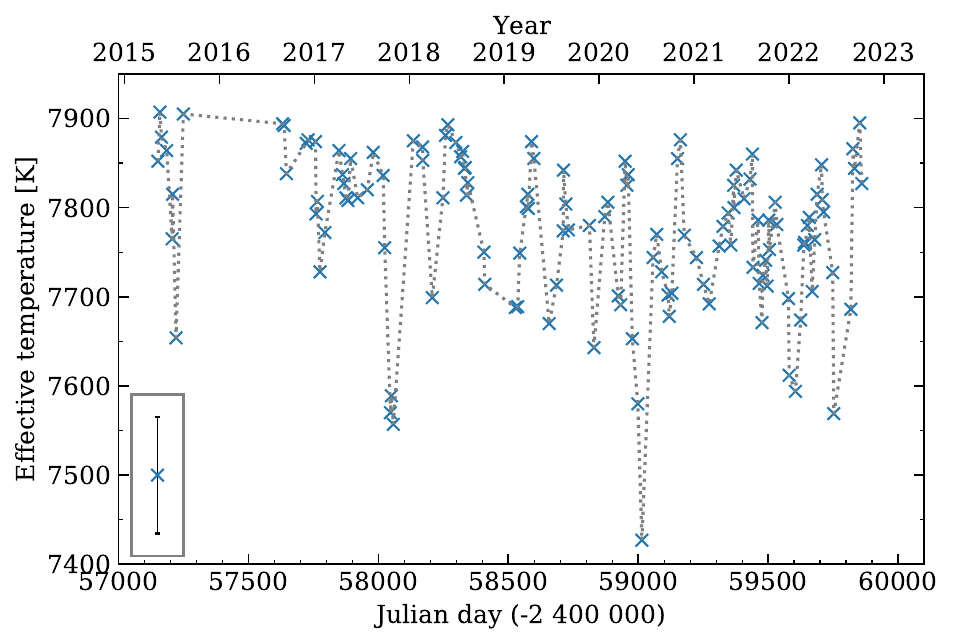}}
    \caption{Effective temperatures of V509\,Cas. In a box on the left, we display the temperature uncertainty derived from LDR measurements.}
    \label{fig:TOteff}
\end{figure}
We determine the effective temperature spectroscopically using line depth ratios (LDR) of temperature sensitive lines. The selection of suitable lines in the case of V509\,Cas and the calibration of the corresponding LDR versus {\teff} is described in Appendix \ref{App:teff}. We used TO long-slit and NOT FIES echelle spectra, which were normalised to the continuum level. In those spectra, we measured the depths of two absorption lines -- \ion{Fe}{II} $\lambda 6446.41$ and \ion{Ca}{I} $\lambda 6439.07$ (wavelength values from NIST\footnote{\url{https://physics.nist.gov/}}), relative to the continuum level. The resulting temperature values are presented in Fig.~\ref{fig:TOteff}. Thanks to the high S/N of the spectra (see Sect.~\ref{section:spectroscopy}), the behaviour of the temperature has been estimated quite convincingly. Based on the results, we can see that over the recent years, the temperature of V509\,Cas has remained at a stable level around 7800~K, with the short-term fluctuations being on average less than 200~K. 

\subsubsection{Comparison with historical values}

\begin{figure}[h]
    \resizebox{\hsize}{!}{\includegraphics{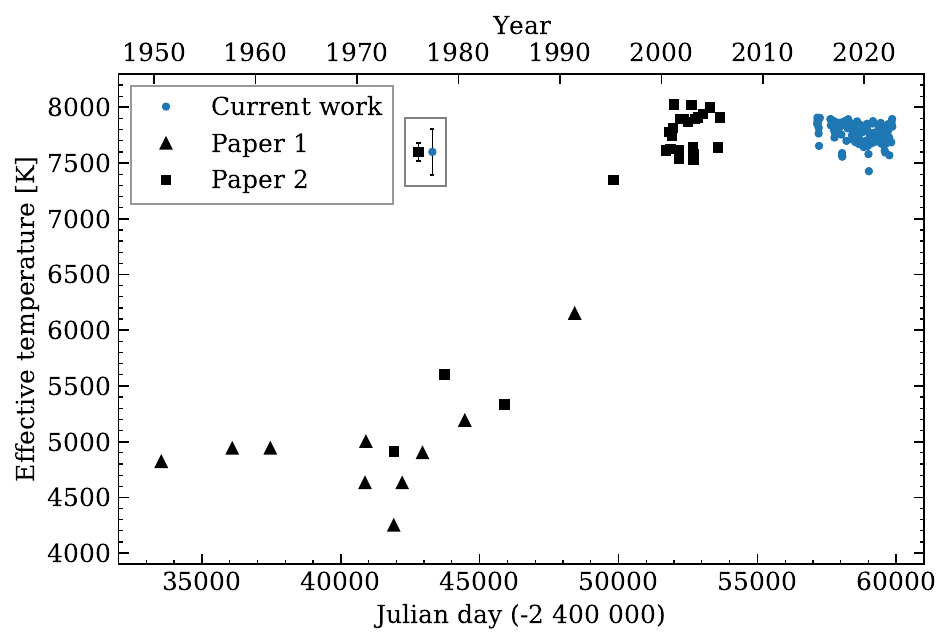}}
    \caption{Variability of \teff of V509\,Cas over the last 70 years, including data from \cite{Nieuwenhuijzen2000} (Paper 1) and \cite{Nieuwenhuijzen2012} (Paper 2). The uncertainties for \cite{Nieuwenhuijzen2012} data near the beginning of 2000s are 60--80~K (sample error bar size is shown in the box), while the uncertainties for earlier data are 72~K in 1973, 120~K in 1978, and 240~K in 1984. For \cite{Nieuwenhuijzen2000}, the uncertainties for all points were not given. }
    \label{fig:teff_historic}
\end{figure}

Figure \ref{fig:teff_historic} illustrates the timeline of temperature variability of V509\,Cas. Since the early 2000s, the temperature of V509\,Cas has remained relatively stable, in stark contrast to the rapid temperature increase in the late 20th century. Comparing the recent temperature values to those found in the beginning of 2000s, we notice a decrease in the amplitude of temperature fluctuations. \citet{Nieuwenhuijzen2012} previously determined effective temperatures by using Kurucz's LTE models. They placed temperatures between 7500--8000~K, with uncertainties of around 80~K. Our results align with theirs, indicating that the star's temperature has settled slightly below 8000~K. In the early 2000s, short-term temperature fluctuations appeared to be more frequent, with temperatures dropping to around 7500~K on several occasions before returning to near-8000~K levels within a month or two. However, during the last seven years of our observations, such occurrences were less common, with just one or two potential temperature drops down to 7500~K. These dips lasted at most 1--2 months before the temperature returned to a higher level. This reduced frequency of significant temperature drops may suggest the stabilisation of the star's atmosphere. 

\subsection{Radial velocity measurements}

\subsubsection{A little bit of history}

Variations in the radial velocity of V509\,Cas are predominantly driven by the star's pulsation cycle. Pulsations are recognised as the primary driver of mass-loss episodes \citep{deJager1998}. Therefore, analysing the radial velocity measurements of spectral lines of metals with different ionisation levels originating from various depths within the star's atmosphere provides insights into the activity in the star's outer layers. These variations may also reflect the passage of shockwaves through the photosphere \citep{Lambert1981}. In 1923 \citeauthor{Harper1923} published radial velocity data for 125 stars, including V509\,Cas (referred to as BD+56 2923), based on observations conducted on 5 epochs between 1921 and 1923. During this period, radial velocity values ranged from $-56.2$ to $-68.2$~km~s$^{-1}$, with a total amplitude of 12~km~s$^{-1}$. The average  radial velocity across these observations was $-60.2~\textrm{km}\, {\textrm{s}}^{-1}$. 

Moving forward to the 1970s, a time marked by mass-loss episodes, the star's atmosphere exhibited significant instability, leading to rapid changes in the observed spectral line profiles. In 1978 \citeauthor{Lambert1978} reported that in September--October 1976, the radial velocity of ionised lines was $-35$~km~s$^{-1}$, while for neutral metallic lines it was $-40$~km~s$^{-1}$. They also noted that the spectral line profiles were asymmetrical with a trailing blue wing. By June 1977, the radial velocities of these metallic lines had shifted to around $-65$~km~s$^{-1}$, a change of $\sim$30~km~s$^{-1}$, and the profiles became almost symmetrical. \citet{Humphreys1978} corroborated these findings and measured star's radial velocity at $-58.3$~km~s$^{-1}$. Based on data from 1979 \citet{Lambert1981} provided radial velocity values for infrared atomic absorption lines (\ion{S}{I}, \ion{Ca}{II} and \ion{Si}{I}) ranging from $-40$ to $-63$~km~s$^{-1}$. Subsequently, in 1987, \citeauthor{Sheffer1987} reported radial velocity data based on the core velocity of many \ion{N}{I} lines near 8700 \AA, with an average value of $-62.5$~km~s$^{-1}$ and a total variability amplitude of 19 km~s$^{-1}$, spanning from maximum $-52$~km~s$^{-1}$ to minimum $-71$ km~s$^{-1}$. More recently, \citet{Klochkova2019a} found radial velocities of various absorption lines within the range of $-52$ to $-71$~km~s$^{-1}$ and a systemic velocity of $-63$~km~s$^{-1}$ for data collected between 1996--2018. The amplitude based on the \ion{N}{I} line was found to be 19~km~s$^{-1}$. The lower part of Fig. \ref{fig:radvel} gives an overview of historic radial velocity measurements. Care should be taken, as the radial velocities reported by different authors are measured from different spectral lines. Nevertheless, the spectral lines of different elements display similar radial velocity variability amplitudes outside eruptive mass-loss periods.

\subsubsection{Our results}\label{sec:radvel}
\begin{figure}[h]
    \resizebox{\hsize}{!}{\includegraphics{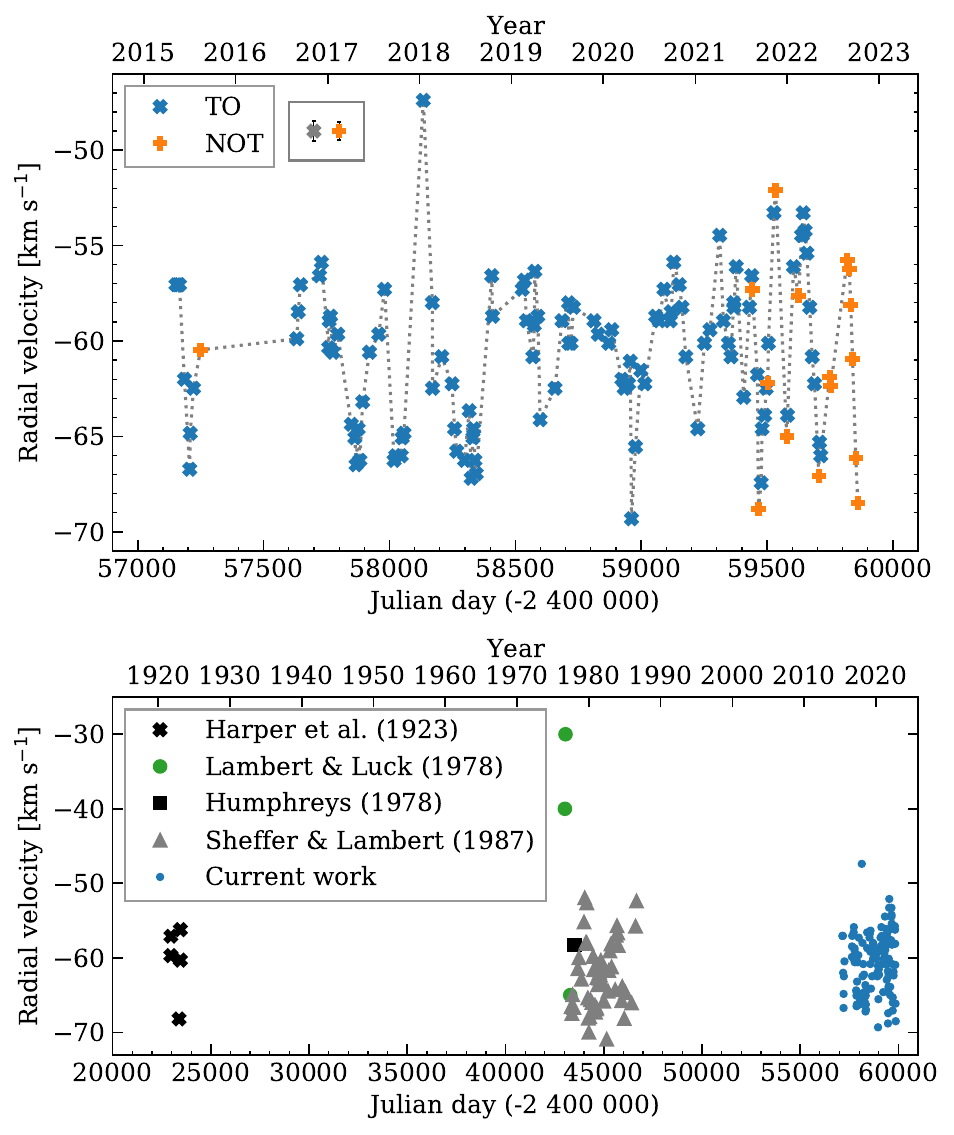}}
    \caption{Radial velocity of V509\,Cas. Upper panel: Our radial velocity measurements based on TO and NOT spectra. Lower panel: Radial velocity measurements from various authors - black crosses correspond to those from \citet{Harper1923}, green circles are for \citet{Lambert1978}, the black square is for \citet{Humphreys1978} (exact observation date not given in the paper, only the year), and grey triangles are for \citet{Lambert1981} (\ion{S}{i}, \ion{Ca}{ii}, and \ion{Si}{i} lines).}
    \label{fig:radvel}
\end{figure}

To characterise the mean expansion and contraction of the stellar surface of V509\,Cas we measure the radial velocity using two relatively strong \ion{Si}{II} absorption lines ($\lambda\lambda 6347.10, 6371.36$). 
These lines are expected to be formed over the full extended atmosphere of the star. The high excitation potential of \ion{Si}{II} ($E_\mathrm{low}$ = 8.12 eV) suggests a contribution from deeper layers, while the significant line depth (up to 50\% of the continuum level, as shown in Fig.~\ref{fig:[CaII]andSiII}) indicates the contribution from outer layers. Additionally, we preferred \ion{Si}{II} lines due to the absence of strongly blending lines and the absence of emission in the line wings, both of which could distort radial velocity measurements. In  Sect. \ref{sec:disc} and in Appendix \ref{App:teff}, we describe the formation and the level of influence of these emission components in more detail.

To improve the accuracy of our radial velocities derived from TO spectra, we performed an additional wavelength scale correction procedure. Using high-resolution FIES spectra with a precise wavelength scale, we determined the location of a nearby diffuse interstellar band (DIB) at $6379.00 \pm 0.01$ \r{A}. This DIB served as a stable reference point, allowing us to calculate the offset value for each TO spectrum, typically of the order of a few hundredths of an angstrom. 

We measured the central wavelength of \ion{Si}{II} lines by fitting a Gaussian function to the central part of the line. Figure~\ref{fig:radvel} shows the radial velocity time series. We assume that the systemic velocity is the average value of all measurements $v_{\mathrm{sys}} =-60.7$~km~s$^{-1}$. Any deviations from this average value indicate dynamics within the star's atmosphere. 

It is astonishing that the radial velocity variability of metal lines has remained at the same level throughout the entire century compared to measurements taken by \citet{Harper1923} in the beginning 1900s, and both before and after the major eruptions in the 1970s, while the star's temperature has increased by several thousand degrees and its brightness has changed by a magnitude. The similarity in the variability of radial velocity of V509 Cas could indicate that the star had been quasi(?)-periodically pulsating before the period of activity in the 1970s. If that's the case, then at the end of the mass-loss episode the star simply restored its previous behaviour.

\section{Interdependence of small-scale variability patterns}

In the previous sections, we demonstrated the disappearance of long-term trends in the variability of V509 Cas's brightness and temperature. This result is consistent with the findings of \citet{Nieuwenhuijzen2012} and \citet{vanGenderen2019}. However, small-scale variability still exists, prompting us to examine the various interdependencies between different parameters in order to provide a more comprehensive description of the current physical state of the star's outer layers.

\begin{figure}[h]
  \resizebox{\hsize}{!}{\includegraphics{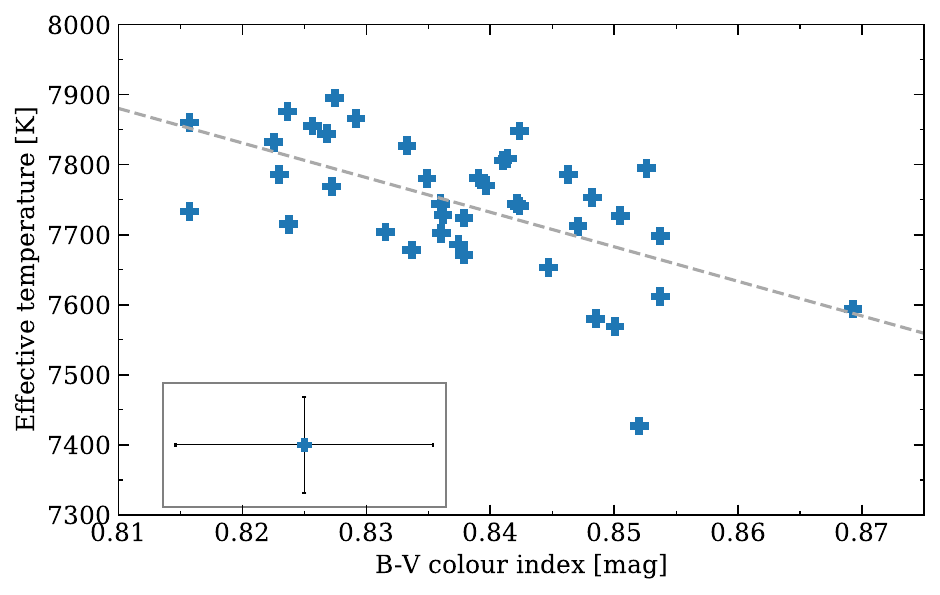}}
  \caption{Effective temperature values based on spectra and $B-V$ colour index from BSM observations on the same epochs.}
  \label{fig:teff_vs_B-V}
\end{figure}

We show the relation between colour and temperature variations in Fig.~\ref{fig:teff_vs_B-V}, using our temperature values derived from spectroscopy and the smoothed $B-V$ colour index values. Even during the current 'calming down' phase of V509 Cas, both the colour and the depth of spectral lines continue to adequately reflect the average effective temperature.

\subsection{Combined timeline}
\begin{figure*}
    \sidecaption
    \includegraphics[width=12cm]{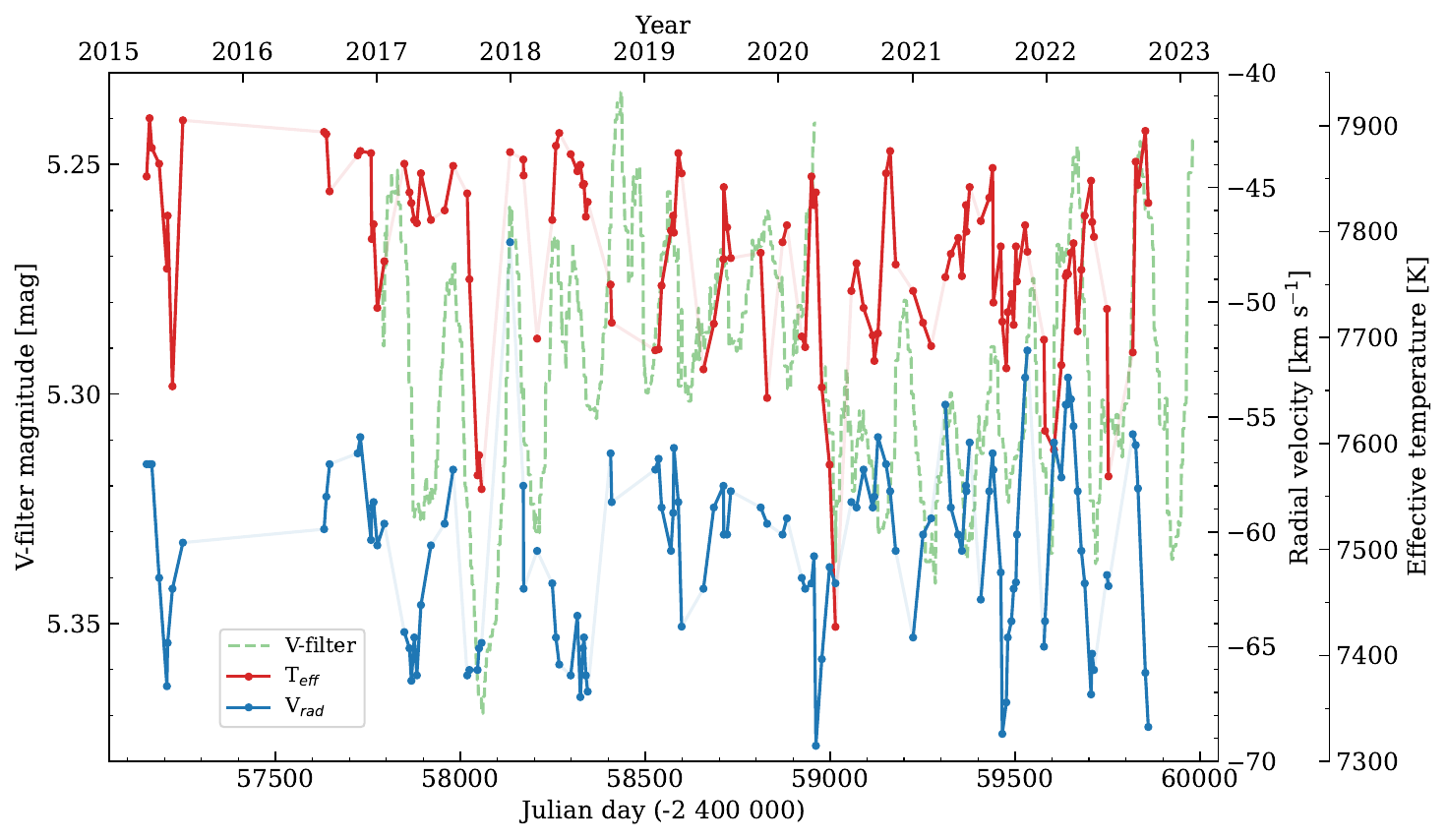}
    \caption{Combined timeline of our measurements: V-filter magnitudes (dashed green line, smoothed \citet{VollmannAAVSO} data from AAVSO), radial velocity measurements (blue), and effective temperatures (red). For the \teff and $v_{\rm rad}$ data, a solid line connects points that have been taken in less than 30-day intervals to further emphasise (dis-)similarities in variability behaviour between different physical parameters. The radial velocity and effective temperature data are included in a table in Appendix \ref{App:radvel_and_teff_data}.}
    \label{fig:combined}
\end{figure*}

In Fig. \ref{fig:combined}, we have drawn three timelines of variability to be able to easily compare the results. Thus, we have an approximately 7-year variability curve to describe the changes in magnitude, temperature and radial velocity of V509 Cas. The variabilities of {\teff} and V-filter magnitudes are qualitatively rather similar. There is a noticeable phase lag between these two and the variability in radial velocity. The lag duration is approximately 30-80 days. Its length is not constant in time and varies with each pulsation period. A similar pattern has been detected in another YHG $\rho$ Cas. Based on observations made in the 1990s and in the beginning of 2000s, the V-filter magnitude curve was found to lag behind the radial velocity curve by approximately 100 days  \citep{Lobel2003}. Connecting only observations taken at intervals of less than 30 days is useful to have a better visual overview of the data. Our time series of NOT spectra taken at 1-week intervals in late 2022 shows smooth radial velocity variability. 

\subsection{Radial velocity of temperature-sensitive lines}

\begin{figure}[h]
    \resizebox{\hsize}{!}{\includegraphics{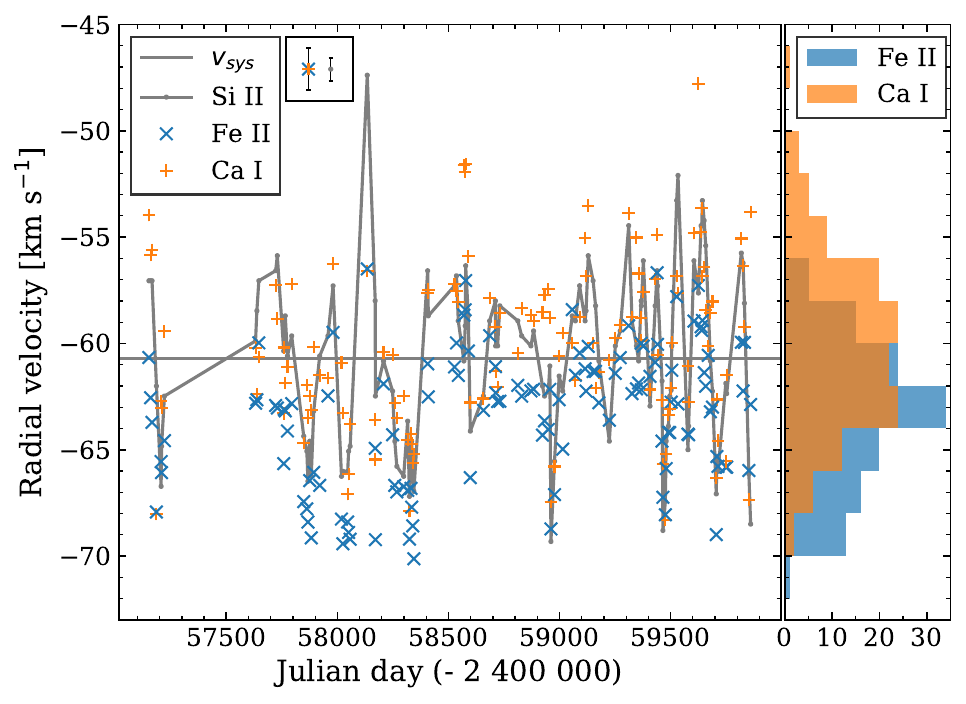}}
    \caption{Centroid radial velocities of \ion{Si}{II}, \ion{Ca}{I}, and \ion{Fe}{II} lines (from \ion{Si}{II} $v_\mathrm{sys}=-60.7$~km\,s$^{-1}$) and their distribution histograms. The error bars for \ion{Ca}{I} and \ion{Fe}{II} (shown in the corner) are larger than the errors for radial velocities from the \ion{Si}{II} line, because the \ion{Fe}{II} and \ion{Ca}{I} lines are much shallower and therefore the results are more affected by noise. Snapshots of both line profiles can be seen in Fig.~\ref{fig:CaIFeIIprofiilid}. }
    \label{fig:CaIFeIIradvel}
\end{figure}

In Sect.~\ref{sec:radvel} we determined the average radial velocity of the stellar disc based on \ion{Si}{II} lines. Now, we investigate the variations in the radial velocities of the spectral lines used for our temperature determination (\ion{Ca}{I} $\lambda6439$ and \ion{Fe}{II} $\lambda6446$). 
Given that both of these lines are asymmetric and relatively weak in comparison to \ion{Si}{II} lines, we could not use the same Gaussian approximation method. Instead, we calculated the centroid ('centre of gravity' of the spectral line) radial velocities for \ion{Fe}{II} and \ion{Ca}{I}. The resulting velocity values and the histograms illustrating their distribution are plotted in Fig.~\ref{fig:CaIFeIIradvel}. 

There is a strong correlation of approximately 0.81 between the radial velocities of \ion{Fe}{II} and \ion{Ca}{I} lines. The position of the \ion{Fe}{II} line centroid oscillates on average at higher blueshift values (by roughly 3.2~km~s$^{-1}$) than the \ion{Si}{II} and \ion{Ca}{I}. This observation calls for further explanation. The spectrum of V509 Cas has a significant number of absorption lines which display emission components above the continuum level in one or both of their wings (c.f. \ion{Sc}{II} which is discussed in Sect. \ref{sec:disc} and the \ion{Fe}{II} $\lambda 7712$ line discussed previously in \cite{Kasikov2022}). The excitation potentials of these lines are generally lower than other lines observed in the spectrum. Circumstellar material affects all lines to a varying degree, depending on excitation energy and abundance of the element. Under such circumstances, we presume that the spectral lines which have no clearly distinguishable emission components are substantially less influenced by the emission from the circumstellar material. Therefore, we have used those ‘clean’ lines (\ion{Ca}{I} $\lambda 6439$, \ion{Fe}{II} $\lambda 64446$ and \ion{Si}{II} $\lambda \lambda 6436,6471$) to estimate the temperature and dynamics of the stellar atmosphere. This does not guarantee that these lines or their wings are entirely undisturbed by the emission - even a small influence, if it exists, can cause the line centroid to shift in the velocity scale. The magnitude of this shift depends on the ratio between the strengths of absorption and emission in each line. From Fig. \ref{fig:ScIIvsradvel} we see that in the case of Sc II the emission is generally stronger in the blue wing of the line, which results in the line absorption centre being shifted more redwards than the average value measured from Si II lines. If the emission influence to the spectral line is very small, or nonexistent, we would see the centroid of the line at lower (or more bluer) velocities - this would explain the on average lower velocity of the \ion{Fe}{II} $\lambda 6446$ line. 

\begin{figure}
    \resizebox{\hsize}{!}{\includegraphics{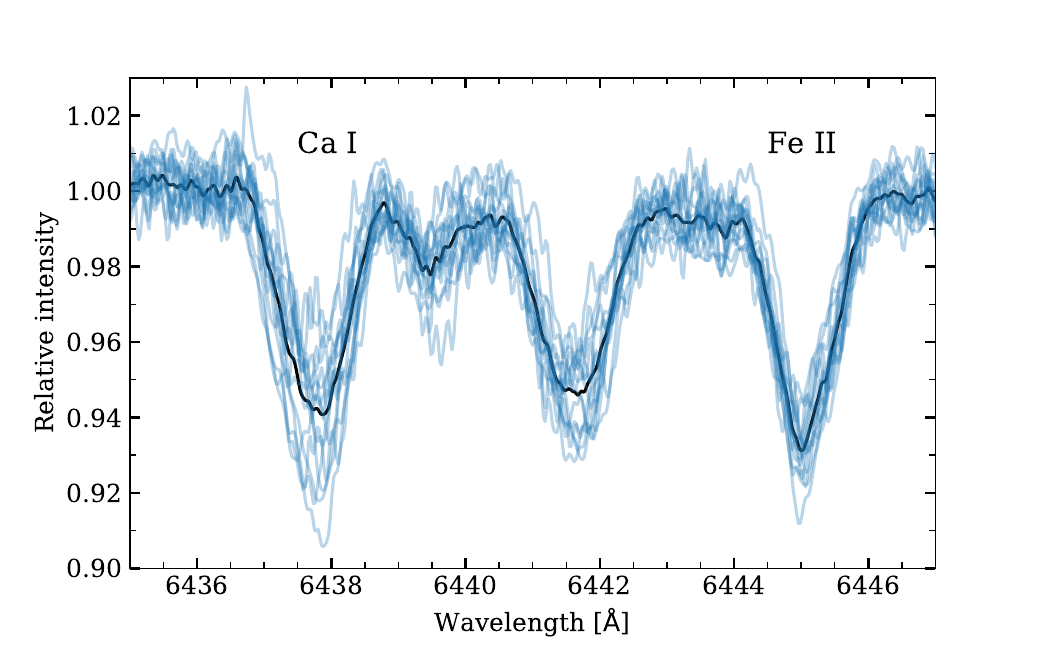}}
    \caption{Variability in the shape of profiles of \ion{Ca}{I} and \ion{Fe}{II} lines that we used for temperature determination. FIES high- and medium-resolution spectra taken at the NOT in 2021 and 2022. The spectra have been superimposed and the dark line traces the shape of the average profile. }
    \label{fig:CaIFeIIprofiilid}
\end{figure}

Figure~\ref{fig:CaIFeIIprofiilid} shows  a series of FIES spectra to illustrate the variability in \ion{Ca}{I} and \ion{Fe}{II} lines. The \ion{Ca}{I} line is more variable in both intensity and shape, while the shape of the \ion{Fe}{II} line is similar on all epochs and its depth is less variable. 

\subsection{Effect of pulsations on the temperature}
\begin{figure*}
\centering
   \includegraphics[width=17cm]{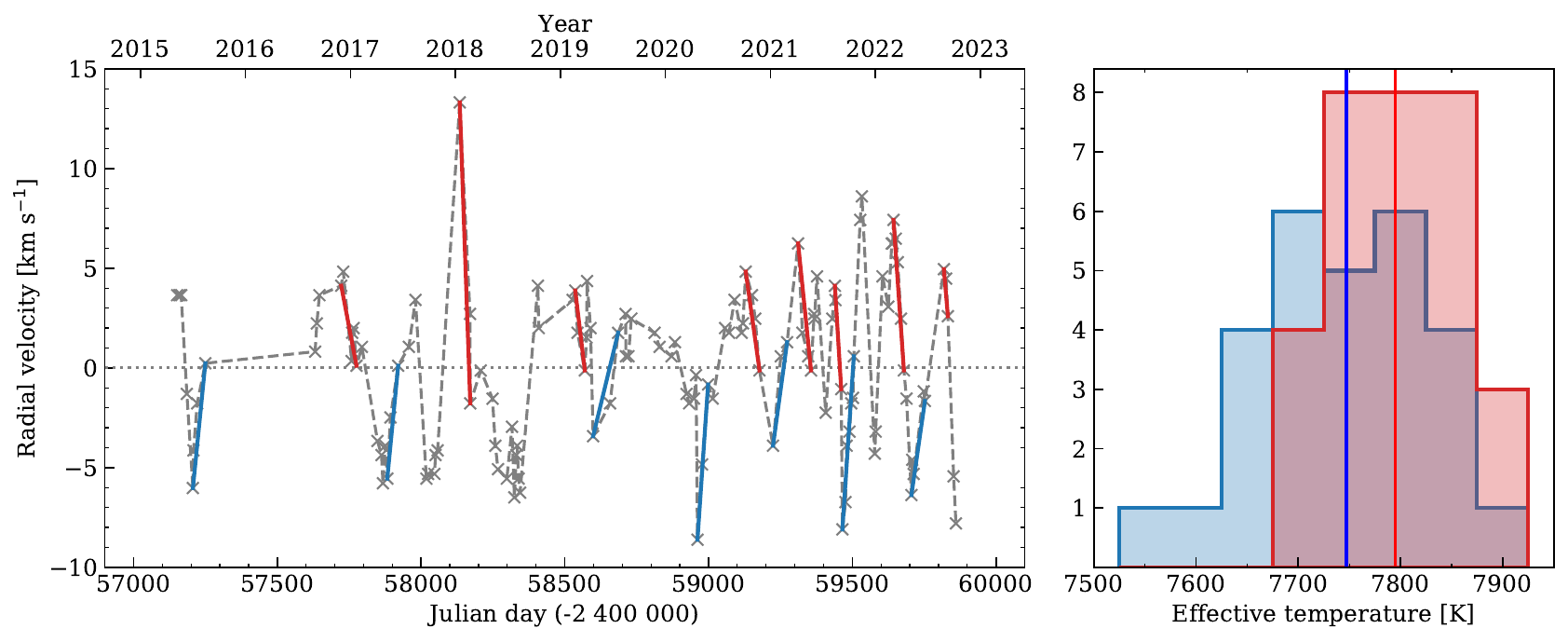}
     \caption{Pulsational behaviour of V509\,Cas. Left: Radial velocity time series of \ion{Si}{II}. As a zero or 'equilibrium' velocity, we have chosen the mean radial velocity of the star at $-60.7$~km~s$^{-1}$ (grey horizontal line). We have marked the epochs, when the stellar disc is at its contraction state (red) and when the star is in an expanding state (blue). We calculated the mean effective temperature for each state. Right: The distribution of mean effective temperatures can be seen on the histogram. During the star's expansion state, the temperatures are on average slightly lower in contrast to the contraction states, when the temperatures are slightly higher.}
     \label{fig:radvel_teff_hist}
\end{figure*}

Radial velocity derived from the \ion{Si}{II} lines could describe the radial component of pulsations of the star -- the mean dynamics across the entire stellar disc. We aim to describe these movements as expansion and contraction phases of the atmosphere of V509\,Cas and assess how the measured radial velocity correlates with the mean temperature level during each stage. 

To achieve this, we identified distinct epochs within our radial velocity time series when the star exhibited either contraction (associated with higher radial velocity values) or expansion (associated with smaller radial velocity values). We assume that both of these states end near the mean (equilibrium) velocity value. Our analysis excludes potential velocity relaxation periods in which representative data was lacking or when the velocity exhibited only small-amplitude variability around the mean. Therefore, on the left side of Fig.~\ref{fig:radvel_teff_hist}, we display these selected 'expansion' and 'contraction' epochs, marked in blue and red, respectively.

We also assume that the 'contraction' and 'expansion' phases are best developed (and have the characteristic temperature) when the corresponding velocities already show  the trend from the local maximum or minimum towards the equilibrium value (epochs of the velocity relaxation). 

We hypothesised that at the end of the contraction stage, the star should reach higher temperatures due to the compression of atmospheric layers, while at the end of the expansion phase, the temperature should decrease due to the drop in the internal energy (enhanced radiative cooling and work done by gas pressure).

The histogram on the right side of Fig.~\ref{fig:radvel_teff_hist} shows the temperatures during these 'expansion' and 'contraction' epochs. We found that the mean effective temperature during the 'contraction' epochs was 7790~K and for the 'expansion' epochs 7740~K. The difference is in the order of magnitude of 1$\sigma$ of the distribution. Both data sets contain temperatures near the overall mean temperature value at 7800 K. However, in the 'expansion' states, there is a higher representation of lower temperatures, while in the 'contraction' state, a number of higher temperatures are found. These findings support our hypothesis, although further observations with better temporal cadence over several pulsational periods would provide a clearer view.

\section{Hints to a disc-like structure}\label{sec:disc}

\begin{figure}[!h]
    \resizebox{\hsize}{!}{\includegraphics{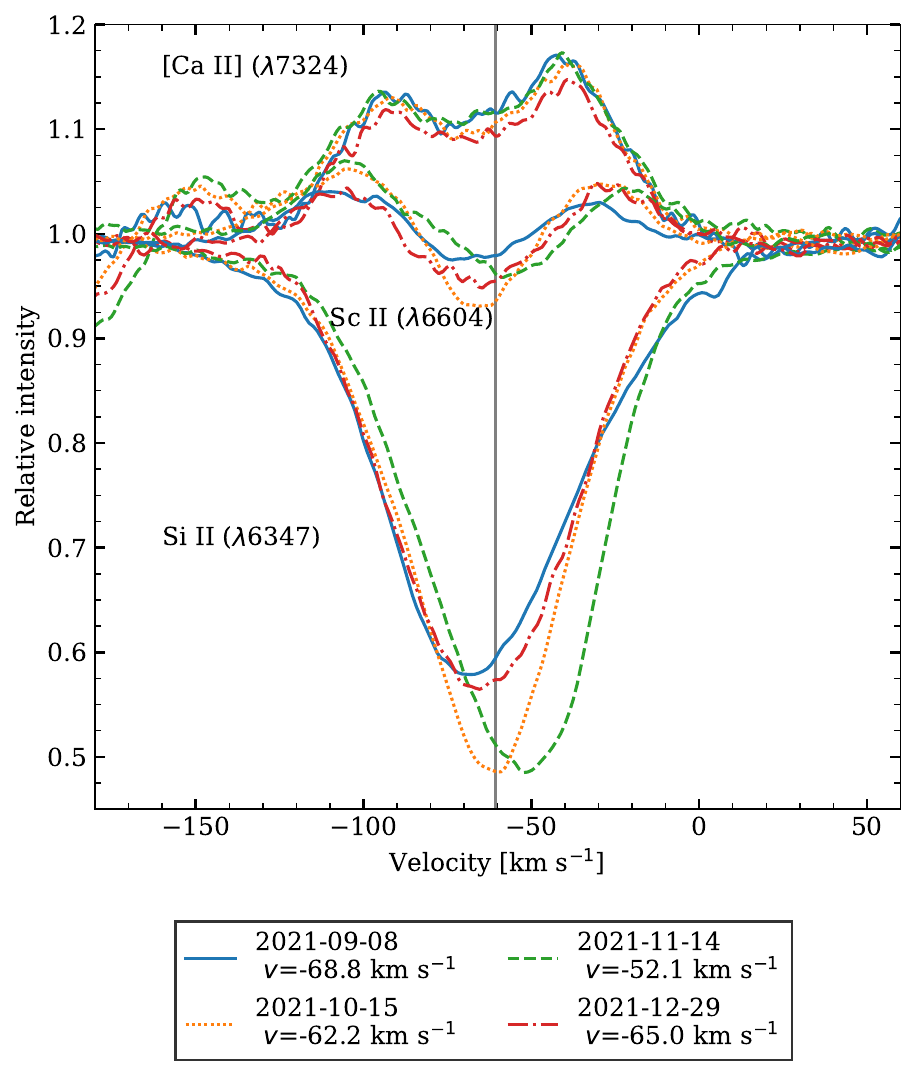}}
    \caption{Comparison of FIES spectra of four epochs over 4 months at the end of 2021 showing three spectral lines: \ion{Si}{II} $\lambda 6347$ \ion{Sc}{II} $\lambda 6604$ and [\ion{Ca}{II}] $\lambda 7324$. \ion{Si}{II} is a strong absorption line and its radial velocity changes with an amplitude of 16~km~s$^{-1}$ over 4 months, while [\ion{Ca}{II}] is a double-peaked emission line that shows barely any changes at the same epochs. The behaviour of \ion{Sc}{II} shows similarities to both of these lines - the position of its emission wings is similar to [\ion{Ca}{II}], but the central absorption shows variability similar to \ion{Si}{II} in both its radial velocity and depth. The grey vertical line marks $v_{\mathrm{sys}}$.}
    \label{fig:[CaII]andSiII}
\end{figure}

A notable feature in the spectra of V509\,Cas is the presence of several forbidden lines ([\ion{Ca}{II}], [\ion{N}{II}], [\ion{O}{I}]) as well as emission components in the wings of absorption lines from various elements (e.g. \ion{Fe}{II}, \ion{Sc}{II}). No large-scale nebulosity has been detected around V509\,Cas \citep{Schuster2003}, thus the existence of these emission features calls for an alternative explanation. 
Similar features have been identified in the spectra of B[e] stars, and it has been suggested that [\ion{Ca}{II}] and [\ion{O}{I}] lines originate from the inner region of the Keplerian gaseous disc surrounding the stars \citep{Kraus2010,Aret2012}. In the case of V509\,Cas, we observe that the [\ion{Ca}{II}] line is double-peaked and remains invariable throughout our observed period. Additionally, [\ion{O}{I}] lines have been found in the spectrum, but unlike the [\ion{Ca}{II}] lines, these lines are single-peaked and rotationally broadened \citep{Aret2017b}. The shape of the [\ion{Ca}{II}] and [\ion{O}{I}] line profiles in our observations implies that we are seeing an inclined rotating disc-like structure around the star. Please note that the usage of the word 'disc' in this context is rather broad, the structure would be most likely akin to a toroidal cloud in Keplerian motion surrounding the star, rather than a thin disc with clear edges.

In Fig.~\ref{fig:[CaII]andSiII}, we show profiles of [\ion{Ca}{II}], \ion{Sc}{II} and [\ion{Si}{II}] at different epochs. The variable \ion{Si}{II} profile illustrates the average movement of the stellar surface and its radial velocity decreases when the star expands and increases when the star contracts. In contrast, the double-peaked [\ion{Ca}{II}] emission line remains virtually unchanged. The \ion{Sc}{II} line shows similarities to both of them, the radial velocity of emission wings does not vary, while the central absorption component varies in sync with the \ion{Si}{II} line in both radial velocity and depth. There are two telluric absorption lines at the edges of [\ion{Ca}{II}] line at approximate velocities -120 and 20 km s$^{-1}$, which have been removed through processing the spectra with a telluric standard star, these lines do not affect the emission peak of [\ion{Ca}{II}]. In \citet{Aret2012}, another [\ion{Ca}{II}] line at $\lambda 7291$ has also been provided as an example; however, that line is strongly blended by telluric features. 

\begin{figure*}
\sidecaption
  \includegraphics[width=12cm]{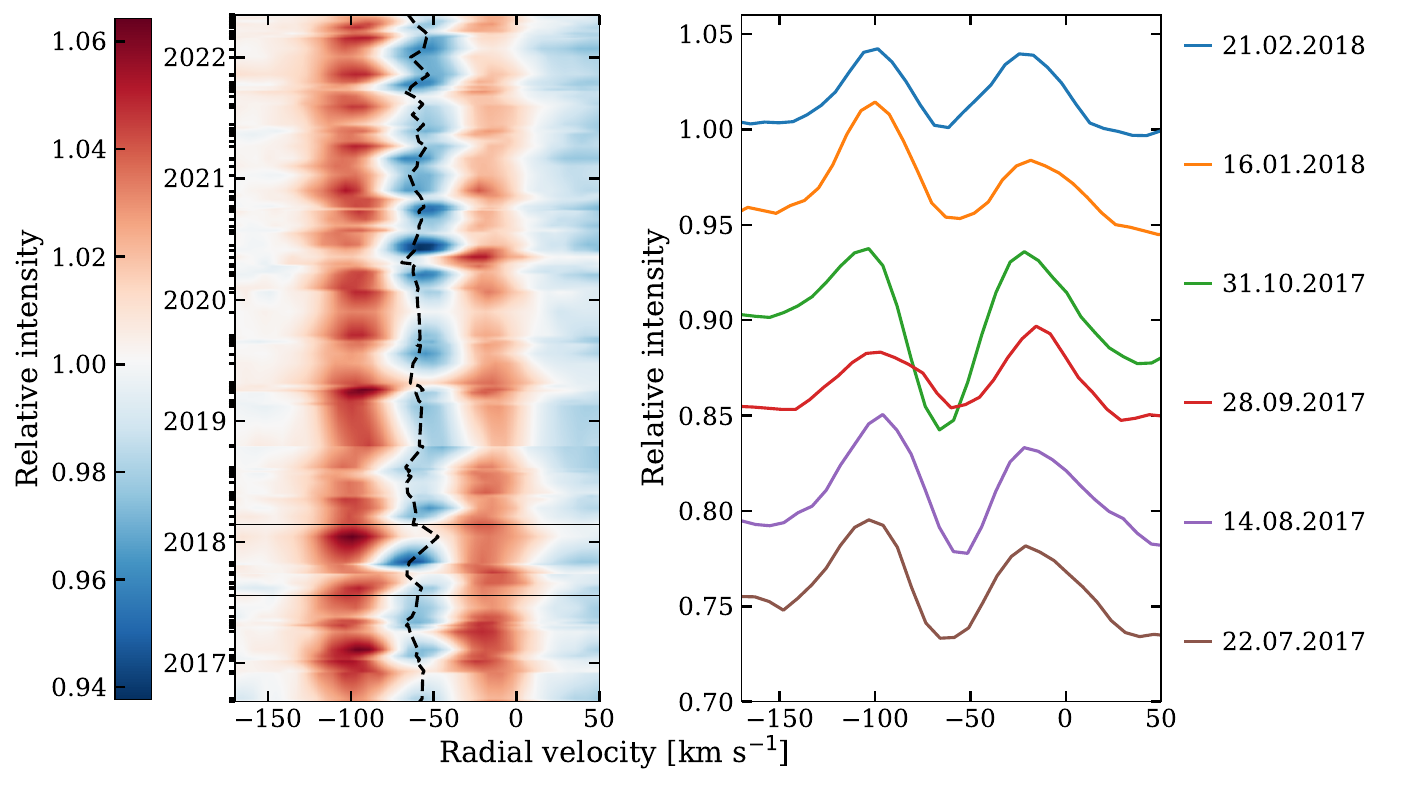}
     \caption{ Variability of \ion{Sc}{II} $\lambda 6604$ line over 6 years. Left: A dynamic spectrum of all TO spectra. White colour corresponds to the continuum level, blue colour indicates absorption and red colour is emission. Darker colour corresponds to stronger absorption or emission strength. Epochs of observations are marked with notches on the left, and gaps between observations have been filled using linear interpolation. The dashed black line shows radial velocity measured from \ion{Si}{II} lines. The box indicates the period of observations seen on the right.
     Right: 6 individual observations in 2017 and 2018. The spectra have been shifted vertically by 0.05~mag between each epoch for clarity. The radial velocity scale on all plots is the same. }
     \label{fig:Sc_heightmap}
\end{figure*}

\begin{figure}
    \resizebox{\hsize}{!}{\includegraphics{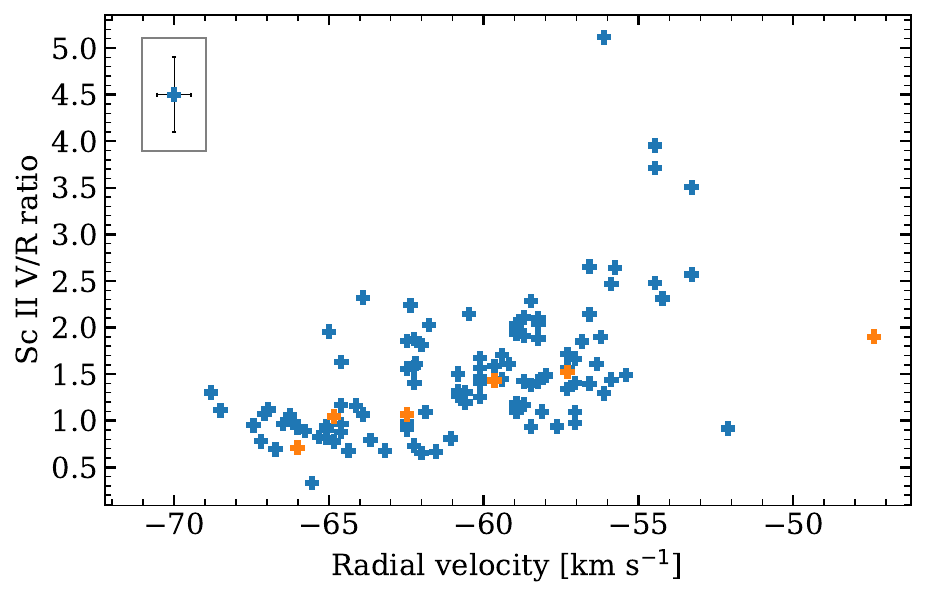}}
    \caption{Radial velocity based on the \ion{Si}{II} lines and the \ion{Sc}{II} violet and red emission component intensity ratio. The epochs displayed on Fig.~\ref{fig:Sc_heightmap} have been highlighted with a lighter orange colour. The sample error in the box shows the size of average error bars, the error size in the y-direction increases proportionally with the calculated ratio.}
   \label{fig:ScIIvsradvel}
\end{figure}

In \cite{Kasikov2022}, we demonstrated that the observed variability of the \ion{Fe}{II} absorption lines with emissions in the wings can be explained by the superposition of absorption and emission components formed in different regions. The absorption component of \ion{Fe}{II} line forms in the pulsating atmosphere and changes in sync with the \ion{Si}{II} line. The double-peaked emission component originates from a circumstellar disc and is not influenced by the pulsations, similar to the [\ion{Ca}{II}] lines. Consequently, in the resulting composite profile, the intensity of the emission components in the blue and red wings of the line is determined by the location and depth of the absorption component, while the emission component remains unchanged.

Similarly to some \ion{Fe}{II} lines, the \ion{Sc}{II} $\lambda 6604$ line has a central absorption component and two emission components on both sides of the central absorption at velocities of roughly $\pm 40$~km~s$^{-1}$. \ion{Sc}{II} line has not received special attention in the literature, but \cite{Harmer1978} observed the spectrum of V509\,Cas in the H$\alpha$ region, among other lines showing the \ion{Sc}{II} line nearby. In 1975 and 1976, at the time when the star underwent an episode of enhanced mass loss, the \ion{Sc}{II} was an absorption line without any emission peaks. 

In Fig. \ref{fig:Sc_heightmap} (left) we have plotted a timeline spanning more than 6 years of \ion{Sc}{II} line variability. The central absorption component is located at roughly $-60$~km~s$^{-1}$. 
With a black dashed line, we have also plotted the radial velocity of the star based on the \ion{Si}{II} line. The radial velocity of the \ion{Sc}{II} absorption behaves synchronously with that of the \ion{Si}{II} line. On the right, we have extracted 6 observation epochs to illustrate the variability of \ion{Sc}{II} at different \ion{Si}{II} radial velocity values. We can see changes in the relative intensity of emission components of \ion{Sc}{II}. At some epochs the violet component is stronger and on others the red one is stronger. The shift occurs on timescales close to a few weeks. The component that is stronger at any point in time is determined by the position of the central absorption, the location of which shifts together with radial velocity as the absorption component 'consumes' either the violet or the red emission wing. 

Overall, there is a correlation between the radial velocity changes and the changes in the intensities of \ion{Sc}{II} line emission components as seen on Fig.~\ref{fig:ScIIvsradvel}. At lower radial velocities, the \ion{Sc}{II} emission components are more likely to have equal intensity or the red component will be slightly stronger, while at higher radial velocities the violet component will dominate. For radial velocities near $v_{\mathrm{sys}}$ both cases are likely. That could be because the change in component intensities is currently ongoing and therefore either one or the other is growing stronger, and it could go either way. Thus, the \ion{Sc}{II} line behaves the same way as \ion{Fe}{II} lines, where the position of enhanced absorption affects the emission components in the line wings.

\section{Conclusions}

We monitored the yellow hypergiant V509\,Cas over the last seven years. We observed no large-scale changes in its variability. The star's \teff has remained stable around 7900~K. In comparison to temperature data from \cite{Nieuwenhuijzen2012}, the amplitude of short-term temperature fluctuations has decreased. 

Based on our radial velocity measurements of \ion{Si}{II} $\lambda\lambda 6347, 6371$ lines, the mean radial velocity over the observed period (systemic velocity) is $v_{\mathrm{sys}}=-60.7$~km~s$^{-1}$ and the total amplitude of the radial velocity variability is up to 20~km~s$^{-1}$. These results are consistent with historical radial velocity data for V509\,Cas. The star displayed similar values almost a century ago, specifically, $-60.2$~km~s$^{-1}$ with an amplitude of 12~km~s$^{-1}$ reported by \cite{Harper1923}. Remarkably, during that period the star had a substantially different temperature ($T_{\mathrm{eff}} \sim 5000$~K) and brightness ($V \sim 5.4$ mag) \citep{Nieuwenhuijzen2012}. 

The star's brightness in the $V$ filter has a short-term variability timescale between 100 and 200 days, also reported by \citet{vanGenderen2019}. In addition to the short-term variability, we have observed a more extended longer-term variability with a timescale of 3--4 years. 

Over the past 50 years, the colour index of V509\,Cas has displayed a gradual shift towards bluer values, with current $B-V$ values at 0.84~mag with a variability amplitude of up to 0.1~mag. This marks a unique phase of stability that has not been previously observed. 

The possible existence of a disc-like structure around the star has been addressed by \citet{Aret2017} based on the [\ion{Ca}{II}] lines in the spectrum. The behaviour of several absorption lines with emission components also supports this hypothesis. We specifically examined the \ion{Sc}{II} $\lambda 6604$ line that has emission components with variable intensities in both wings at approximately $\pm40$~km~s$^{-1}$ from the central absorption. While the radial velocity of the absorption component behaves similarly to the \ion{Si}{II} line due to pulsations, the radial velocity of the emission components does not change, and their intensity is affected by the current state of the absorption.

We conclude that YHG V509\,Cas currently resides in the hotter half of the yellow void (\citet{vanGenderen2019}, Fig. 1), exhibiting relatively mild variability. To determine its further evolutionary path, continuous monitoring with a sufficient cadence is necessary.

\begin{acknowledgements}
    The authors would like to thank the observers of Nordic Optical Telescope, Tartu Observatory, AAVSO and Bright Star Monitoring programme for the data used in this paper. \\
    
    This work has made use of the ground-based research infrastructure of Tartu Observatory, funded through the projects TT8 (Estonian Research Council) and KosEST (EU Regional Development Fund).\\

     This study is based on observations made with the Nordic Optical Telescope, owned in collaboration by the University of Turku and Aarhus University, and operated jointly by Aarhus University, the University of Turku and the University of Oslo, representing Denmark, Finland and Norway, the University of Iceland and Stockholm University at the Observatorio del Roque de los Muchachos, La Palma, Spain, of the Instituto de Astrofisica de Canarias.\\
     
     We acknowledge with thanks the variable star observations from the AAVSO International Database contributed by observers worldwide and used in this research.\\

     This paper includes data collected with the TESS mission, obtained from the MAST data archive at the Space Telescope Science Institute (STScI). Funding for the TESS mission is provided by the NASA Explorer Program. STScI is operated by the Association of Universities for Research in Astronomy, Inc., under NASA contract NAS 5–26555.\\

     This work has made use of data from the European Space Agency (ESA) mission
    {\it Gaia} (\url{https://www.cosmos.esa.int/gaia}), processed by the {\it Gaia}
    Data Processing and Analysis Consortium (DPAC,
    \url{https://www.cosmos.esa.int/web/gaia/dpac/consortium}). Funding for the DPAC
    has been provided by national institutions, in particular the institutions
    participating in the {\it Gaia} Multilateral Agreement.\\
     
     This project has received funding from the European Union’s Framework Programme for Research and Innovation Horizon 2020 under the Marie Skłodowska-Curie Grant Agreement No. 823734.

     This work was supported by the Estonian Research Council grant PRG 2159.
\end{acknowledgements}

\bibliographystyle{aa}
\bibliography{references.bib}

\begin{appendix}

\section{Temperatures using the line-depth ratio (LDR) method}\label{App:teff}

Spectral line-depth ratios as temperature indicators were suggested in pioneering papers by \cite{Gray1991}
and \cite{Gray1994}. 
\cite{Kovtyukh2007} 
demonstrated that this approach gives precise effective temperatures for FGK supergiants. Guided by these papers, the LDR method has been used recently for \object{$\rho$~Cas} (a yellow hypergiant in the Galaxy) by \cite{Kraus2019} 
and for evolved hypergiants in the Magellanic Clouds by \cite{Kourniotis2022}. 
The last two papers have noticed the necessity to select specific 
pairs of spectral lines for the LDR method in the particular cases
of studied hypergiants.

In the case of V509\,Cas the selection of proper lines for LDR is 
constrained by the fact that its absorption lines are quite often 
disturbed with the circumstellar emission components. This 
phenomenon was first noticed in the middle of 1970s when 
V509\,Cas experienced the period of main activity (e.g. \cite{Lambert1978}),
and is observable until now in the spectra used in this work. The temporal changes are in the prevalence of the ionisation state most influenced by the emission -- first it was neutral atoms, but now singly ionised ones.
It has been noticed (cf. e.g. \cite{Fukue2015} 
and \cite{Matsunaga2021})  
that there are other more common constraints limiting the precision of the empirical LDR 
calibration. The selected pairs should belong to one chemical element
to avoid abundance effects, and the LDR \textasciitilde \teff relationship depends on the surface gravity and the metallicity of the calibrators.

\begin{figure}
    \resizebox{\hsize}{!}{\includegraphics{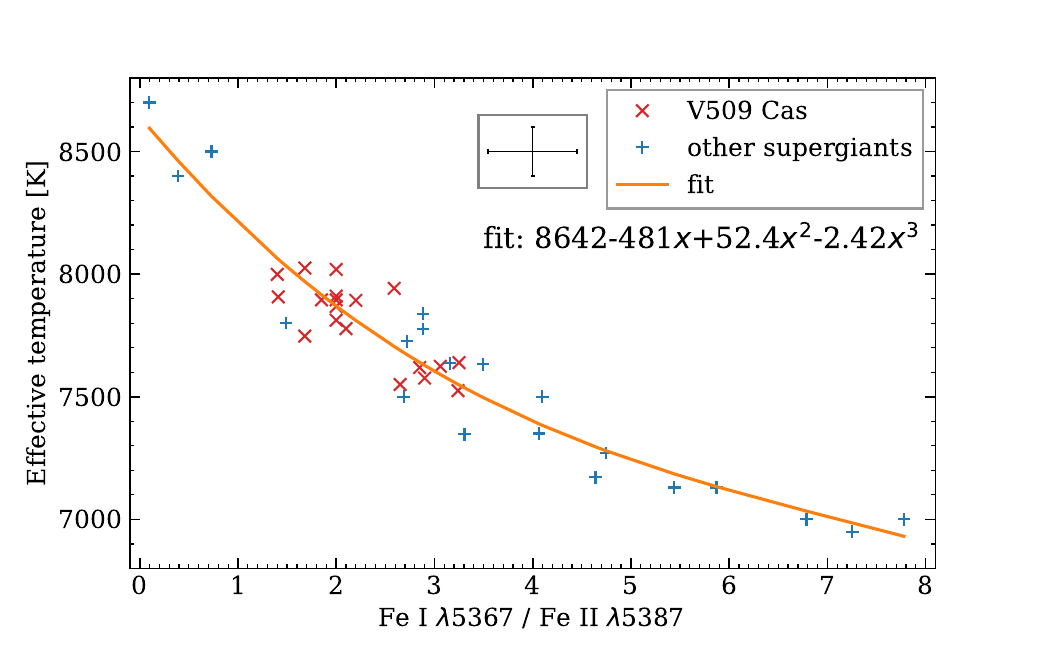}}
    \caption{Temperature fit of V509\,Cas and additional calibration stars based on \ion{Fe}{I} $\lambda$5367.47 and \ion{Fe}{II} $\lambda$5387.06 lines.}
    \label{fig:TeffFeII_lines}
\end{figure}

We aimed to take these limitations into account and decided to 
rely on the LDR of \ion{Fe}{I} and \ion{Fe}{II} lines. By visual inspection of V509\,Cas spectra we have noticed that \ion{Fe}{I} lines with E$_{\mathrm{low}}$ >~ 4.0~eV and
\ion{Fe}{II} lines with E$_\mathrm{low}$ >$\sim$ 5.0~eV have no clearly distinguishable emission components and are consequently very minimally or not at all influenced by CS emission.
In addition, our intention has been to have the new 
LDR $\sim$ \teff relationship comparable with the temperature scale for V509\,Cas obtained in \cite{Nieuwenhuijzen2012}. 
Therefore, the absorption lines \ion{Fe}{I} $\lambda 5367.47$ (E$_{\mathrm{low}}$ = 4.42~eV) and 
\ion{Fe}{II} $\lambda 5387.06$ (E$_{\mathrm{low}}$ = 10.52~eV) were selected which have been measurable 
both in spectra used in \cite{Nieuwenhuijzen2012} 
and in the spectra of additional calibrators. As these, we selected the supergiants 
from the list presented by \cite{Kovtyukh2007} 
in Table 4, and added a few more well-analysed supergiants to cover the temperature range $7000\ldots 8500$~K. We list all the calibrator stars in the Table~\ref{tab:FeIIcalibspectra} indicating the sources of their high resolution spectra and of their 
temperatures applied for calibration. The obtained correlation 
between LDR and temperature is depicted in Fig.~\ref{fig:TeffFeII_lines} including its
third order polynomial approximation.

\begin{figure}
    \resizebox{\hsize}{!}{\includegraphics{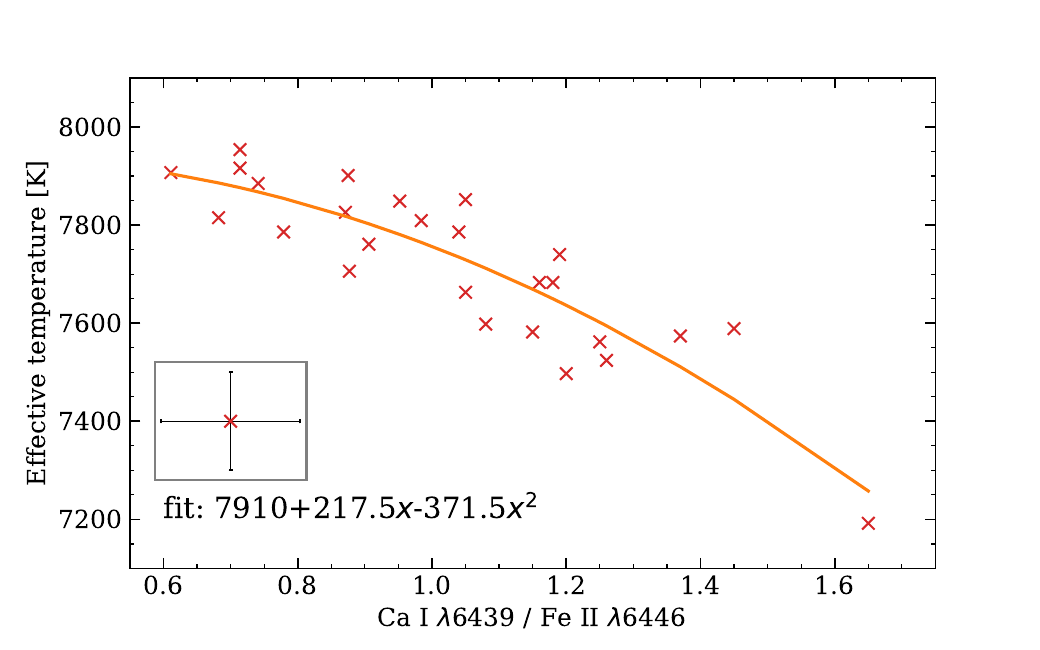}}
    \caption{Effective temperatures of V509\,Cas and the corresponding \ion{Ca}{I}~$\lambda$6439.08 / Fe~II~$\lambda$6446.41 line depth ratio. }
    \label{fig:TeffCaIFeIIlines}
\end{figure}

Taking into account that our spectral monitoring of V509\,Cas 
includes many epochs when the mentioned green lines were not 
observed (the red interval commonly used is $\sim6300 \ldots 6600$\AA), we decided to transform this calibration into the second one which is based solely on the high resolution spectra of 
V509\,Cas that cover both the green and red regions. These 
archival and new spectra are listed in the Table~\ref{tab:CaIFeIIcalib} where 
we include the temperatures estimated with the 'green' 
calibration. In the mentioned red region there are several 
non-blended \ion{Fe}{I} and \ion{Fe}{II} lines which would be suitable for 
LDR calibration (e.g. \cite{Lambert1978} 
indicate the sensitivity of lines \ion{Fe}{I} $\lambda 6430.85$ and \ion{Fe}{II} $\lambda 6432.68$ in the case of V509\,Cas during the period of its activity in the middle of 1970s). However, in contemporary spectra, the CS  emission components limit our choice. By visual inspection, we have noticed that the depth of the \ion{Ca}{I} $\lambda 6439.08$ line (E$_{\mathrm{low}}$=2.53~eV) 
is remarkably variable while the depth of \ion{Fe}{II} $\lambda 6446.41$ line 
(E$_{\mathrm{low}}$=6.22~eV) changes much less.
Both of these lines are free of distinguishable emission components and are minimally  influenced by CS emission. 
Measuring this LDR in the spectra listed in Table~\ref{tab:CaIFeIIcalib}, we obtained 
the relationship demonstrated in Fig.~\ref{fig:TeffCaIFeIIlines}. The corresponding 
second order polynomial fit has been used for our monitoring 
of the V509\,Cas surface temperature. It should be noticed that
the values estimated using the (two) correlative relationship(s)
have quite remarkable formal errors (up to 140~K) but they are
suitable to convert the variable LDR measured with high 
precision into the temperature scale.

\begin{table}[h]
 \caption[]{\label{tab:FeIIcalibspectra}Spectra used for temperature calibration based on the ratio of Fe~$\lambda\lambda 5367,5387$ lines (r5367/5387).}
\begin{tabular}{lcc}
 \hline \hline
 Object & $T_\mathrm{eff}$ & r5367/5387
 \\ \hline
 HD 197345& 8700 $^{(e)}$& 0.099 $^{(1)}$\\
 HD 213470& 8400 $^{(e)}$& 0.387 $^{(2)}$\\
 HD 210221& 8500 $^{(d)}$& 0.731 $^{(1)}$\\
 V509\,Cas&  7999 $^{(g)}$& 1.400 $^{(6)}$\\
 V509\,Cas&  7907 $^{(g)}$& 1.410 $^{(6)}$\\
 HD 164514& 7800  $^{(f)}$& 1.491 $^{(1)}$\\
 V509\,Cas&  8025 $^{(g)}$& 1.680 $^{(6)}$\\
 V509\,Cas&  7747 $^{(g)}$& 1.680 $^{(6)}$\\
 V509\,Cas&  7895 $^{(g)}$& 1.850 $^{(6)}$\\
 V509\,Cas&  7895 $^{(g)}$& 2.000 $^{(6)}$\\
 V509\,Cas&  8019 $^{(g)}$& 2.000 $^{(6)}$\\
 V509\,Cas&  7910 $^{(g)}$& 2.000 $^{(6)}$\\
 V509\,Cas&  7868 $^{(g)}$& 2.000 $^{(6)}$\\
 V509\,Cas&  7812 $^{(g)}$& 2.000 $^{(6)}$\\
 V509\,Cas&  7778 $^{(g)}$& 2.100 $^{(6)}$\\
 V509\,Cas&  7893 $^{(g)}$& 2.200 $^{(6)}$\\
 V509\,Cas&  7942 $^{(g)}$& 2.590 $^{(6)}$\\
 V509\,Cas&  7550 $^{(g)}$& 2.650 $^{(6)}$\\
 HD 85123&  7500 $^{(c)}$& 2.687 $^{(3)}$\\
 HD 34578&  7726 $^{(f)}$& 2.719 $^{(1)}$\\
 V509\,Cas&  7619 $^{(g)}$& 2.850 $^{(6)}$\\ 
 HD 74180&  7839 $^{(a)}$& 2.882 $^{(1)}$\\
 HD 68601&  7776 $^{(f)}$& 2.885 $^{(1)}$\\
 V509\,Cas&  7576 $^{(g)}$& 2.900 $^{(6)}$\\
 V509\,Cas&  7624 $^{(g)}$& 3.060 $^{(6)}$\\
 HD 1457& 7636$^{(f)}$& 3.156 $^{(2)}$\\
 V509\,Cas&  7525 $^{(g)}$& 3.240 $^{(6)}$\\
 V509\,Cas&  7639 $^{(g)}$& 3.250 $^{(6)}$\\
 HD 7927& 7347$^{(a)}$& 3.306 $^{(2)}$\\
 HD 9167& 7632$^{(f)}$& 3.492 $^{(2)}$\\
 HD 179821& 7350 $^{(b)}$& 4.065 $^{(4)}$\\
 HD 80404&  7500 $^{(c)}$& 4.098 $^{(5)}$\\
 HD 173638& 7171 $^{(a)}$& 4.639 $^{(2)}$\\
 HD 36673&  7270 $^{(a)}$& 4.750 $^{(1)}$\\
 HD 75276&  7130 $^{(a)}$& 5.442 $^{(5)}$\\
 HD 25291&  7130 $^{(a)}$& 5.871 $^{(1)}$\\
 HD 194951& 7000 $^{(a)}$& 6.788 $^{(1)}$\\
 HD 61227&  6950 $^{(a)}$& 7.255 $^{(2)}$\\
 HD 220102& 7000 $^{(a)}$& 7.780 $^{(2)}$\\
\hline
\end{tabular}
\tablebib{(a)~\citet{Luck2014};
(b) \citet{Sahin2016}; 
(c) \citet{Neiner2017}; 
(d) \citet{Martin2018}; 
(e) \citet{Firnstein2012}; 
(f) \cite{Kovtyukh2007}; 
(g) \citet{Nieuwenhuijzen2012}. 

Archives of spectra: (1)  PolarBase; (2)  Elodie; (3)  ESO FEROS; (4)  ESO X-SHOOTER; (5)  ESO UVES; (6)  Tartu observatory.
}
\end{table}

\begin{table}[h]
 \caption[]{\label{tab:CaIFeIIcalib}V509\,Cas spectra used for temperature calibration based on the ratio of \ion{Ca}{I}~$\lambda$6439 and \ion{Fe}{II}~$\lambda$6446 lines.}
\begin{tabular}{lcc}
 \hline \hline
Date & \ion{Ca}{I}/\ion{Fe}{II} & $T_{\mathrm{eff}}$
 \\ \hline
30.09.2022 $^{(a)}$  & 0.611  & 7907\\
03.09.2022 $^{(a)}$  & 0.682  & 7815\\
10.09.2022 $^{(a)}$  & 0.714  & 7916\\
15.08.2015 $^{(a)}$  & 0.714  & 7954\\
27.05.2002 $^{(e)}$  & 0.741  & 7885\\
05.08.2015 $^{(a)}$  & 0.779  & 7786\\
05.05.2022 $^{(a)}$  & 0.871  & 7826\\
07.10.2022 $^{(a)}$  & 0.875  & 7901\\
02.10.2001 $^{(c)}$  & 0.877  & 7706\\
22.07.2005 $^{(d)}$  & 0.906  & 7761\\
20.08.2009 $^{(a)}$  & 0.952  & 7849\\
14.11.2021 $^{(a)}$  & 0.984  & 7809\\
15.08.2021 $^{(a)}$  & 1.040  & 7786\\
09.07.2000 $^{(c)}$  & 1.050  & 7852\\
15.10.2021 $^{(a)}$  & 1.050  & 7663\\
18.06.2022 $^{(a)}$  & 1.080  & 7598\\
29.12.2021 $^{(a)}$  & 1.150  & 7582\\
13.06.2011 $^{(f)}$   &1.160   & 7683\\
05.10.1996 $^{(b)}$  & 1.180  & 7683\\
27.08.2022 $^{(a)}$  & 1.190  & 7740\\
02.05.1996 $^{(g)}$  & 1.200  & 7497\\
17.04.1995 $^{(c)}$  & 1.250  & 7562\\
14.02.2022 $^{(a)}$  & 1.260  & 7524\\
08.09.2021 $^{(a)}$  & 1.370  & 7574\\
22.06.2022 $^{(a)}$  & 1.450  & 7589\\
03.07.1996 $^{(g)}$  & 1.650  & 7192\\
\hline
\end{tabular}
\tablebib{(a)  NOT FIES; (b)  Elodie; (c)  WHT UES; (d)  NOT SOFIN (A. Lobel); (e)  TNG (G. Israelian); (f)  PolarBase; (g)  SAO Lynx echelle spectrograph (V. Klochkova)
}
\end{table}

\clearpage
\section{Observations log}\label{app:observations}
\begin{table}[h]
\caption{Spectroscopic observations with FIES at the NOT. The resolution and S/N of the spectra are given.}   
\centering
\begin{tabular}{l c c c}
\hline\hline
Date & Observer & Resolution & S/N\\ 
\hline
    2015-08-14 & Kuutma & high & 300\\
    2021-08-14 & Kasikov, Knudstrup & medium & 200\\
    2021-09-07 & Keniger & medium & 350\\
    2021-10-14 & Kasikov & medium & 300\\
    2021-11-14 & Kasikov & medium & 300\\
    2021-12-29 & Kasikov & medium & 230\\
    2022-02-14 & Paraskeva & medium & 200\\ 
    2022-05-05 & Kasikov & medium & 320\\ 
    2022-06-17 & Pinter & medium & 380\\ 
    2022-06-21 & Keniger & medium & 320\\ 
    2022-08-26 & Matilainen & high & 230\\
    2022-09-03 & Dürfeldt Pedros, Biancalani & high & 210\\
    2022-09-09 & Thomsen & high & 200\\
    2022-09-16 & Keniger & high & 100\\ 
    2022-09-29 & Matilainen & high & 190\\
    2022-10-07 & Thomsen & high & 190\\
\hline
\end{tabular}
\label{table:FIESobs}
\end{table}

\begin{table}\label{table:TOobs}
\caption{Observations made with TO 1.5 m telescope AZT-12.}
\centering
\begin{tabular}{ll|ll}
\hline
\hline
Date & Observer & Date & Observer\\
\hline
2015-05-08 & A. Aret                                & 2019-08-15 & A. Aret                    \\
2015-05-16 & A. Aret                                & 2019-08-16 & A. Aret                    \\
2015-05-22 & A. Aret                                & 2019-08-25 & Aret, Checha               \\
2015-06-11 & T. Eenm\"ae                            & 2019-09-04 & V. Checha                  \\
2015-07-02 & A. Aret                                & 2019-11-24 & V. Checha                  \\
2015-07-04 & A. Aret                                & 2019-12-11 & T. Eenm\"ae                \\
2015-07-17 & A. Aret                                & 2020-01-22 & A. Aret                    \\
2016-08-30 & A. Aret                                & 2020-02-03 & A. Kasikov                 \\
2016-09-05 & A. Aret                                & 2020-03-14 & A. Aret                    \\
2016-09-14 & Aret, Kivila                           & 2020-03-23 & A. Kasikov                 \\
2016-11-29 & A. Aret                                & 2020-04-09 & T. Eenm\"ae                \\
2016-12-06 & A. Aret                                & 2020-04-16 & T. Eenm\"ae                \\
2017-01-04 & A. Aret                                & 2020-04-21 & T. Eenm\"ae                \\
2017-01-06 & A. Aret                                & 2020-05-07 & T. Eenm\"ae                \\
2017-01-11 & A. Aret                                & 2020-05-28 & A. Aret                    \\
2017-01-21 & T. Eenm\"ae                            & 2020-06-13 & T. Eenm\"ae                \\
2017-02-09 & T. Eenm\"ae                            & 2020-07-26 & A. Aret                    \\
2017-04-04 & \"U. Kivila                            & 2020-08-09 & A. Aret                    \\
2017-04-18 & A. Aret                                & 2020-08-28 & A. Aret                    \\
2017-04-23 & A. Aret                                & 2020-09-22 & A. Kasikov                 \\
2017-05-01 & A. Aret                                & 2020-09-26 & A. Aret                    \\
2017-05-08 & A. Aret                                & 2020-10-06 & A. Kasikov                 \\
2017-05-19 & A. Aret                                & 2020-10-28 & V. Checha                  \\
2017-06-15 & T. Eenm\"ae                            & 2020-11-08 & V. Checha                  \\
2017-07-22 & A. Aret                                & 2020-11-23 & T. Eenm\"ae                \\
2017-08-14 & A. Aret                                & 2021-01-09 & V. Checha                  \\
2017-09-22 & \"U. Kivila                            & 2021-02-05 & V. Checha                  \\
2017-09-28 & A. Aret                                & 2021-02-27 & V. Checha                  \\
2017-10-20 & T. Eenm\"ae                            & 2021-04-06 & A. Aret                    \\
2017-10-24 & T. Eenm\"ae                            & 2021-04-21 & V. Checha                  \\
2017-10-31 & T. Eenm\"ae                            & 2021-05-11 & A. Kasikov                 \\
2018-01-16 & A. Aret                                & 2021-05-21 & V. Checha                  \\
2018-02-20 & A. Aret                                & 2021-06-01 & V. Checha                  \\
2018-02-21 & A. Aret                                & 2021-06-02 & A. Kasikov                 \\
2018-03-29 & \"U. Kivila                            & 2021-06-11 & A. Aret                    \\
2018-05-09 & Kasikov, Eenm\"ae                      & 2021-07-11 & T. Eenm\"ae                \\
2018-05-19 & A. Aret                                & 2021-08-03 & V. Checha                  \\
2018-05-28 & A. Aret                                & 2021-08-12 & V. Checha                  \\
2018-06-28 & A. Aret                                & 2021-09-03 & V. Checha                  \\
2018-07-16 & A. Aret                                & 2021-09-18 & V. Checha                  \\
2018-07-24 & A. Aret                                & 2021-09-22 & V. Checha                  \\
2018-07-31 & A. Aret                                & 2021-10-02 & V. Checha                  \\
2018-08-03 & A. Aret                                & 2021-10-08 & V. Checha                  \\
2018-08-08 & A. Aret                                & 2021-10-17 & H. Ramler                  \\
2018-08-13 & A. Kasikov                             & 2021-11-08 & V. Checha                  \\
2018-10-14 & A. Aret                                & 2022-01-01 & V. Checha                  \\
2018-10-17 & A. Aret                                & 2022-01-25 & V. Checha                  \\
2019-02-12 & A. Aret                                & 2022-02-26 & V. Checha                  \\
2019-02-21 & Aret, Kasikov                          & 2022-03-01 & V. Checha                  \\
2019-03-01 & A. Aret                                & 2022-03-04 & V. Checha                  \\
2019-03-26 & K. Metsoja                             & 2022-03-12 & V. Checha                  \\
2019-04-01 & A. Kasikov                             & 2022-03-19 & T. Eenm\"ae                \\
2019-04-03 & A. Kasikov                             & 2022-03-30 & A. Aret                    \\
2019-04-15 & A. Kasikov                             & 2022-04-09 & V. Mitrokhina              \\
2019-04-24 & A. Aret                                & 2022-04-18 & V. Mitrokhina              \\
2019-06-22 & A. Aret                                & 2022-05-08 & V. Checha                  \\
2019-07-20 & A. Aret                                & 2022-05-13 & V. Mitrokhina              \\
\hline
\end{tabular}
\end{table}

\clearpage
\onecolumn

\section{Radial velocity and effective temperature data}

\setlength{\LTcapwidth}{\textwidth}
\begin{longtable}[h]{lllll|lllll}
\caption{\label{App:radvel_and_teff_data} Observation dates, Julian days (JD) and measured radial velocity ($v$) and effective temperature (\teff) results, including information about the observatory and telescope used TO - Tartu Observatory 1.5 m telescope, NOT - Nordic Optical Telescope. Mean radial velocity errors of Gaussian fitting of \ion{Si}{II} lines are 0.54 km s$^{-1}$ for TO data and 0.47 km s$^{-1}$ for NOT data.}\\
\hline\hline
Date & JD & $v$ (km s$^{-1}$) & \teff (K) & Obs. & Date & JD & $v$ (km s$^{-1}$) & \teff (K) & Obs. \\
\hline
\endfirsthead
\caption{continued.}\\
\hline\hline
Date & JD & $v$ (km s$^{-1}$) & \teff (K) & Obs. & Date & JD & $v$ (km s$^{-1}$)& \teff (K) & Obs. \\
\hline
\endhead
\hline
\endfoot
2015-05-08 & 2457151.544 & -57.05 & 7852$\pm$53 & TO & 2020-02-03 & 2458883.199 & -59.41 & 7806$\pm$60 & TO \\
2015-05-16 & 2457159.501 & -57.05 & 7907$\pm$29 & TO & 2020-03-14 & 2458923.297 & -62.01 & 7701$\pm$82 & TO \\
2015-05-22 & 2457165.463 & -57.05 & 7879$\pm$40 & TO & 2020-03-23 & 2458932.552 & -62.48 & 7691$\pm$98 & TO \\
2015-06-11 & 2457185.407 & -62.01 & 7864$\pm$46 & TO & 2020-04-09 & 2458949.51 & -62.24 & 7852$\pm$45 & TO \\
2015-07-02 & 2457206.409 & -66.72 & 7765$\pm$62 & TO & 2020-04-16 & 2458956.533 & -61.06 & 7825$\pm$59 & TO \\
2015-07-04 & 2457208.41 & -64.84 & 7815$\pm$49 & TO & 2020-04-21 & 2458961.466 & -69.32 & 7837$\pm$53 & TO \\
2015-07-17 & 2457221.398 & -62.48 & 7654$\pm$92 & TO & 2020-05-07 & 2458977.452 & -65.54 & 7653$\pm$117 & TO \\
2015-08-14 & 2457249.597 & -60.47 & 7905$\pm$18 & NOT & 2020-05-28 & 2458998.476 & -61.54 & 7580$\pm$108 & TO \\
2016-08-30 & 2457631.532 & -59.88 & 7894$\pm$31 & TO & 2020-06-13 & 2459014.433 & -62.24 & 7427$\pm$159 & TO \\
2016-09-05 & 2457637.514 & -58.47 & 7892$\pm$32 & TO & 2020-07-26 & 2459057.39 & -58.70 & 7744$\pm$80 & TO \\
2016-09-14 & 2457646.354 & -57.05 & 7838$\pm$58 & TO & 2020-08-09 & 2459071.424 & -58.94 & 7770$\pm$72 & TO \\
2016-11-29 & 2457722.251 & -56.58 & 7872$\pm$42 & TO & 2020-08-28 & 2459090.41 & -57.29 & 7728$\pm$ & 77 \\
2016-12-06 & 2457729.507 & -55.87 & 7876$\pm$43 & TO & 2020-09-22 & 2459115.299 & -58.94 & 7702$\pm$70 & TO \\
2017-01-04 & 2457758.423 & -60.36 & 7874$\pm$36 & TO & 2020-09-26 & 2459119.618 & -58.47 & 7678$\pm$82 & TO \\
2017-01-06 & 2457760.35 & -58.94 & 7793$\pm$65 & TO & 2020-10-06 & 2459129.375 & -55.87 & 7704$\pm$76 & TO \\
2017-01-11 & 2457765.355 & -58.70 & 7807$\pm$59 & TO & 2020-10-28 & 2459151.265 & -57.05 & 7855$\pm$44 & TO \\
2017-01-21 & 2457775.364 & -60.59 & 7728$\pm$77 & TO & 2020-11-08 & 2459162.353 & -58.23 & 7876$\pm$35 & TO \\
2017-02-09 & 2457794.306 & -59.65 & 7772$\pm$58 & TO & 2020-11-23 & 2459177.255 & -60.83 & 7769$\pm$66 & TO \\
2017-04-04 & 2457848.446 & -64.37 & 7864$\pm$40 & TO & 2021-01-09 & 2459224.327 & -64.60 & 7744$\pm$75 & TO \\
2017-04-18 & 2457862.453 & -65.07 & 7837$\pm$43 & TO & 2021-02-05 & 2459251.225 & -60.12 & 7714$\pm$85 & TO \\
2017-04-23 & 2457867.435 & -66.49 & 7827$\pm$51 & TO & 2021-02-27 & 2459273.29 & -59.41 & 7692$\pm$86 & TO \\
2017-05-01 & 2457875.446 & -64.60 & 7811$\pm$56 & TO & 2021-04-06 & 2459311.585 & -54.46 & 7757$\pm$73 & TO \\
2017-05-08 & 2457882.411 & -66.25 & 7808$\pm$56 & TO & 2021-04-21 & 2459326.516 & -58.94 & 7779$\pm$68 & TO \\
2017-05-19 & 2457893.415 & -63.19 & 7855$\pm$38 & TO & 2021-05-11 & 2459346.406 & -60.12 & 7794$\pm$61 & TO \\
2017-06-15 & 2457920.444 & -60.59 & 7811$\pm$55 & TO & 2021-05-21 & 2459356.421 & -60.83 & 7758$\pm$69 & TO \\
2017-07-22 & 2457957.508 & -59.65 & 7820$\pm$52 & TO & 2021-06-01 & 2459367.413 & -58.00 & 7825$\pm$52 & TO \\
2017-08-14 & 2457980.527 & -57.29 & 7862$\pm$41 & TO & 2021-06-02 & 2459368.467 & -58.23 & 7800$\pm$58 & TO \\
2017-09-22 & 2458018.5 & 66.25 & 7836$\pm$54 & TO & 2021-06-11 & 2459377.393 & -56.11 & 7842$\pm$47 & TO \\
2017-09-28 & 2458024.5 & 66.02 & 7755$\pm$95 & TO & 2021-07-11 & 2459407.416 & -62.95 & 7810$\pm$53 & TO \\
2017-10-20 & 2458046.5 & 66.02 & 7570$\pm$127 & TO & 2021-08-03 & 2459430.46 & -58.23 & 7832$\pm$56 & TO \\
2017-10-24 & 2458050.5 & 65.07 & 7589$\pm$110 & TO & 2021-08-12 & 2459439.467 & -56.58 & 7860$\pm$45 & TO \\
2017-10-31 & 2458057.5 & 64.84 & 7557$\pm$108 & TO & 2021-08-14 & 2459441.524 & -57.30 & 7733$\pm$70 & NOT \\
2018-01-16 & 2458134.5 & 47.39 & 7875$\pm$43 & TO & 2021-09-03 & 2459461.53 & -61.77 & 7786$\pm$70 & TO \\
2018-02-20 & 2458170.255 & -57.99 & 7868$\pm$43 & TO & 2021-09-07 & 2459465.606 & -68.80 & 7715$\pm$64 & NOT \\
2018-02-21 & 2458171.128 & -62.48 & 7853$\pm$49 & TO & 2021-09-18 & 2459476.519 & -67.43 & 7671$\pm$102 & TO \\
2018-03-29 & 2458207.607 & -60.83 & 7699$\pm$87 & TO & 2021-09-22 & 2459480.346 & -64.60 & 7724$\pm$80 & TO \\
2018-05-09 & 2458248.495 & -62.24 & 7811$\pm$60 & TO & 2021-10-02 & 2459490.349 & -63.89 & 7741$\pm$70 & TO \\
2018-05-19 & 2458258.44 & -64.60 & 7881$\pm$37 & TO & 2021-10-08 & 2459496.322 & -62.48 & 7712$\pm$76 & TO \\
2018-05-28 & 2458267.447 & -65.78 & 7893$\pm$31 & TO & 2021-10-14 & 2459502.565 & -62.20 & 7786$\pm$37 & NOT \\
2018-06-28 & 2458298.426 & -66.25 & 7873$\pm$38 & TO & 2021-10-17 & 2459505.314 & -60.12 & 7753$\pm$68 & TO \\
2018-07-16 & 2458316.443 & -63.66 & 7857$\pm$41 & TO & 2021-11-08 & 2459527.364 & -53.28 & 7806$\pm$56 & TO \\
2018-07-24 & 2458324.504 & -67.20 & 7863$\pm$38 & TO & 2021-11-14 & 2459533.454 & -52.10 & 7781$\pm$48 & NOT \\
2018-07-31 & 2458331.376 & -65.07 & 7844$\pm$47 & TO & 2021-12-29 & 2459578.347 & -65.00 & 7698$\pm$63 & NOT \\
2018-08-03 & 2458334.366 & -64.60 & 7845$\pm$43 & TO & 2022-01-01 & 2459581.188 & -63.89 & 7612$\pm$101 & TO \\
2018-08-08 & 2458339.4 & -66.25 & 7814$\pm$55 & TO & 2022-01-25 & 2459605.227 & -56.11 & 7594$\pm$105 & TO \\
2018-08-13 & 2458344.354 & -66.96 & 7828$\pm$52 & TO & 2022-02-14 & 2459625.331 & -57.63 & 7674$\pm$69 & NOT \\
2018-10-14 & 2458406.378 & -56.58 & 7750$\pm$78 & TO & 2022-02-26 & 2459637.298 & -54.46 & 7758$\pm$68 & TO \\
2018-10-17 & 2458409.364 & -58.70 & 7714$\pm$89 & TO & 2022-03-01 & 2459640.288 & -54.46 & 7761$\pm$69 & TO \\
2019-02-12 & 2458527.279 & -57.29 & 7688$\pm$96 & TO & 2022-03-04 & 2459643.305 & -53.28 & 7760$\pm$75 & TO \\
2019-02-21 & 2458536.21 & -56.82 & 7689$\pm$98 & TO & 2022-03-12 & 2459651.289 & -54.22 & 7780$\pm$70 & TO \\
2019-03-01 & 2458544.222 & -58.94 & 7749$\pm$78 & TO & 2022-03-19 & 2459658.239 & -55.40 & 7789$\pm$64 & TO \\
2019-03-26 & 2458569.596 & -60.83 & 7801$\pm$55 & TO & 2022-03-30 & 2459669.539 & -58.23 & 7706$\pm$91 & TO \\
2019-04-01 & 2458575.557 & -59.18 & 7815$\pm$57 & TO & 2022-04-09 & 2459679.473 & -60.83 & 7764$\pm$72 & TO \\
2019-04-03 & 2458577.542 & -56.35 & 7799$\pm$64 & TO & 2022-04-18 & 2459688.576 & -62.24 & 7815$\pm$59 & TO \\
2019-04-15 & 2458589.456 & -58.70 & 7874$\pm$44 & TO & 2022-05-05 & 2459705.708 & -67.08 & 7848$\pm$36 & NOT \\
2019-04-24 & 2458598.456 & -64.13 & 7855$\pm$51 & TO & 2022-05-08 & 2459708.437 & -65.31 & 7809$\pm$60 & TO \\
2019-06-22 & 2458657.43 & -62.48 & 7670$\pm$99 & TO & 2022-05-13 & 2459713.519 & -66.02 & 7795$\pm$62 & TO \\
2019-07-20 & 2458685.471 & -58.94 & 7713$\pm$79 & TO & 2022-06-17 & 2459748.673 & -61.88 & 7727$\pm$53 & NOT \\
2019-08-15 & 2458711.517 & -58.00 & 7774$\pm$72 & TO & 2022-06-21 & 2459752.7 & -62.36 & 7569$\pm$84 & NOT \\
2019-08-16 & 2458712.462 & -60.12 & 7842$\pm$45 & TO & 2022-08-26 & 2459818.538 & -55.75 & 7686$\pm$68 & NOT \\
2019-08-25 & 2458721.368 & -60.12 & 7804$\pm$54 & TO & 2022-09-03 & 2459826.464 & -56.22 & 7866$\pm$28 & NOT \\
2019-09-04 & 2458731.349 & -58.23 & 7775$\pm$64 & TO & 2022-09-09 & 2459832.547 & -58.11 & 7844$\pm$32 & NOT \\
2019-11-24 & 2458812.245 & -58.94 & 7780$\pm$80 & TO & 2022-09-16 & 2459839.555 & -60.94 & - \footnote{The missing $T_{\mathrm{eff}}$ value on 2022-09-16 is due to low S/N of the spectrum, this data point has not been included in Fig. \ref{fig:combined} or Fig. \ref{fig:radvel_teff_hist}, but the radial velocity is included in Fig. \ref{fig:radvel}.} & NOT \\
2019-12-11 & 2458829.583 & -59.65 & 7643$\pm$102 & TO & 2022-09-29 & 2459852.543 & -66.14 & 7895$\pm$23 & NOT \\
2020-01-22 & 2458871.315 & -60.12 & 7790$\pm$64 & TO & 2022-10-07 & 2459860.467 & -68.50 & 7827$\pm$44 & NOT \\

\end{longtable}

\end{appendix}

\end{document}